\documentclass[lettersize,journal]{IEEEtran}
\usepackage{amsmath,amsfonts}
\usepackage[linesnumbered,ruled,vlined]{algorithm2e}
\usepackage{pifont}
\usepackage{amsthm}
\usepackage{algpseudocode}
\usepackage{tcolorbox}
\usepackage{array}
\usepackage[caption=false,font=normalsize,labelfont=sf,textfont=sf]{subfig}
\usepackage{textcomp}
\usepackage{stfloats}
\usepackage{url}
\usepackage{verbatim}
\usepackage{graphicx}
\usepackage{ulem}
\usepackage{amsfonts}
\usepackage[utf8]{inputenc}
\usepackage{setspace}
\usepackage{cleveref}
\crefname{section}{§}{§§}
\Crefname{section}{§}{§§}
\usepackage{caption}
\usepackage{booktabs}
\usepackage{makecell}
\usepackage{MnSymbol}
\usepackage{utfsym}
\usepackage{filecontents, pgffor}
\usepackage{balance}
\usepackage{latexsym}
\usepackage{color,xcolor,colortbl}
\usepackage{multirow}

\definecolor{mySourceNodecolor}{HTML}{ddf1e8}
\definecolor{myOtherNodecolor}{HTML}{385775}
 \usepackage{tikz}

\usepackage{listings, xcolor}
\definecolor{verylightgray}{rgb}{.97,.97,.97}

\lstdefinelanguage{Solidity}{
	keywords=[1]{anonymous, assembly, assert, balance, break, call, callcode, case, catch, class, constant, continue, constructor, contract, debugger, default, delegatecall, delete, do, else, emit, event, experimental, export, external, false, finally, for, function, gas, if, implements, import, in, indexed, instanceof, interface, internal, is, length, library, log0, log1, log2, log3, log4, memory, modifier, new, payable, pragma, private, protected, public, pure, push, require, return, returns, revert, selfdestruct, send, solidity, storage, struct, suicide, super, switch, then, this, throw, transfer, true, try, typeof, using, value, view, while, with, addmod, ecrecover, keccak256, mulmod, ripemd160, sha256, sha3}, 
	keywordstyle=[1]\color{blue}\bfseries,
	keywords=[2]{address, bool, byte, bytes, bytes1, bytes2, bytes3, bytes4, bytes5, bytes6, bytes7, bytes8, bytes9, bytes10, bytes11, bytes12, bytes13, bytes14, bytes15, bytes16, bytes17, bytes18, bytes19, bytes20, bytes21, bytes22, bytes23, bytes24, bytes25, bytes26, bytes27, bytes28, bytes29, bytes30, bytes31, bytes32, enum, int, int8, int16, int24, int32, int40, int48, int56, int64, int72, int80, int88, int96, int104, int112, int120, int128, int136, int144, int152, int160, int168, int176, int184, int192, int200, int208, int216, int224, int232, int240, int248, int256, mapping, string, uint, uint8, uint16, uint24, uint32, uint40, uint48, uint56, uint64, uint72, uint80, uint88, uint96, uint104, uint112, uint120, uint128, uint136, uint144, uint152, uint160, uint168, uint176, uint184, uint192, uint200, uint208, uint216, uint224, uint232, uint240, uint248, uint256, var, void, ether, finney, szabo, wei, days, hours, minutes, seconds, weeks, years},	
	keywordstyle=[2]\color{teal}\bfseries,
	keywords=[3]{block, blockhash, coinbase, difficulty, gaslimit, number, timestamp, msg, data, gas, sender, sig, value, now, tx, gasprice, origin},	
	keywordstyle=[3]\color{violet}\bfseries,
	identifierstyle=\color{black},
	sensitive=true,
	comment=[l]{//},
	morecomment=[s]{/*}{*/},
	commentstyle=\color{gray}\ttfamily,
	stringstyle=\color{red}\ttfamily,
	morestring=[b]',
	morestring=[b]"
}

\begin{document}
\title{ScamSweeper: Detecting Illegal Accounts in Web3 Scams via Transactions Analysis}

\author{Xiaoqi Li, Wenkai Li, Zhijie Liu, Meikang Qiu, Zhiquan Liu, Sen Nie, Zongwei Li, Shi Wu, Yuqing Zhang
\thanks{Xiaoqi Li, Wenkai Li, and Zongwei Li are with Hainan University, Haikou, 570228, China (e-mail: csxqli@ieee.org, cswkli@hainanu.edu.cn, lizw1017@hainanu.edu.cn)}
\thanks{Zhijie Liu is with Ohio State University, Columbus, OH, USA (e-mail: liu.13062@osu.edu)}
\thanks{Meikang Qiu is with the College of Computer and Cyber Science, Augusta University, Augusta, GA, USA (e-mail: qiumeikang@yahoo.com)}
\thanks{Zhiquan Liu is with the Jinan University, Guangzhou, 510632, China (e-mail: zqliu@jnu.edu.cn)}
\thanks{Sen Nie and Shi Wu are with the Keen Security Lab, Tencent, Shanghai, China (e-mail: \{snie, shiwu\}@tencent.com)}
\thanks{Yuqing Zhang is with the University of Chinese Academy of Sciences, Beijing, 100049, China (e-mail: zhangyq@nipc.org.cn)}
\thanks{Wenkai Li (cswkli@hainanu.edu.cn) is the corresponding author.}
\thanks{This manuscript is an extended version of our work~\cite{ase2024poster}. It has been extended more than 80\% over the ASE conference version, including: (1) Optimization of the contribution and motivation of the paper (Sec. I \& Sec. III).
(2) Elaboration on the detailed principles of the framework (Sec. IV).
(3) Enhancement of the analysis of the experiments (Sec. V).
(4) Addition of discussion with the scams detected by our framework (Sec. VI).
 }
}

\markboth{IEEE Transactions on Information Forensics and Security,~Vol.~XX, No.~X, XXX~2025}%
{Li \MakeLowercase{\textit{et al.}}: Detecting Illegal Accounts in Web3 Scams via Transactions Analysis}
\newcommand{\blackding}[1]{\ding{\numexpr181+#1\relax}}
\newcommand{\whiteding}[1]{\ding{\numexpr171+#1\relax}}

\maketitle

\begin{abstract}
The web3 applications have recently been growing, especially on the Ethereum platform, starting to become the target of scammers. The web3 scams, imitating the services provided by legitimate platforms, mimic regular activity to deceive users. However, previous studies have primarily concentrated on de-anonymization and phishing nodes, neglecting the distinctive features of web3 scams. Moreover, the current phishing account detection tools utilize graph learning or sampling algorithms to obtain graph features. However, large-scale transaction networks with temporal attributes conform to a power-law distribution, posing challenges in detecting web3 scams. 
To overcome these challenges, we present ScamSweeper, a \textit{novel} framework that emphasizes the dynamic evolution of transaction graphs, to identify web3 scams on Ethereum. ScamSweeper samples the network with a structure temporal random walk, which is an optimized sample walking method that considers both temporal attributes and structural information. Then, the directed graph encoder generates the features of each subgraph during different temporal intervals, sorting as a sequence. Moreover, a variational Transformer is utilized to extract the dynamic evolution in the subgraph sequence. Furthermore, we collect a large-scale transaction dataset consisting of web3 scams, phishing, and normal accounts, which are from the first 18 million block heights on Ethereum. Subsequently, we comprehensively analyze the distinctions in various attributes, including nodes, edges, and degree distribution. Our experiments indicate that ScamSweeper outperforms SIEGE, Ethident, and PDTGA in detecting web3 scams, achieving a weighted F1-score improvement of at least 17.29\% with the base value of 0.59. In addition, ScamSweeper in phishing node detection achieves at least a 17.5\% improvement over DGTSG and BERT4ETH in F1-score from 0.80.

\end{abstract}

\begin{IEEEkeywords}
Web3 scam, Deep learning, Malicious account, Transaction analysis
\end{IEEEkeywords}

\section{Introduction}
\label{sec:intro}

Detecting malicious accounts on the blockchain networks~\cite{huang2025blockchain,yu2024blockchain} is a significant area of research~\cite{kim2024drainclog,ratra2024graph}. Due to its high transaction volume and anonymity, the Ethereum blockchain~\cite{huang2025blockchain,li2025atomgraph} has been a prime target for fraudulent activities~\cite{li2025penetrating}. Numerous studies have focused on identifying phishing accounts to safeguard users from scams~\cite{grover2016node2vec,he2023txphishscope, li2023siege,li2022ttagn, perozzi2014deepwalk, poursafaei2021sigtran, wu2023understanding,  zhou2022behavior}. However, with the emergence of Ethereum, web3 scams have also emerged, where certain services engage in covert malicious activities within the blockchain~\cite{li2025atomgraph}. For example, Venom Drainer~\cite{li2025facial} offered web3 services that generated over \$27 million in profits from about 15,000 victims~\cite{VenomDrainers2023}. These scams attract users and generate profits by concealing specific malicious activities within legitimate services, such as concealing scams in NFT airdrops~\cite{airdrop2024NFT,li2025ckg,li2025beyond}.

\begin{figure}[ht]
\vspace{-2ex}
\centering
\includegraphics[width=\linewidth]{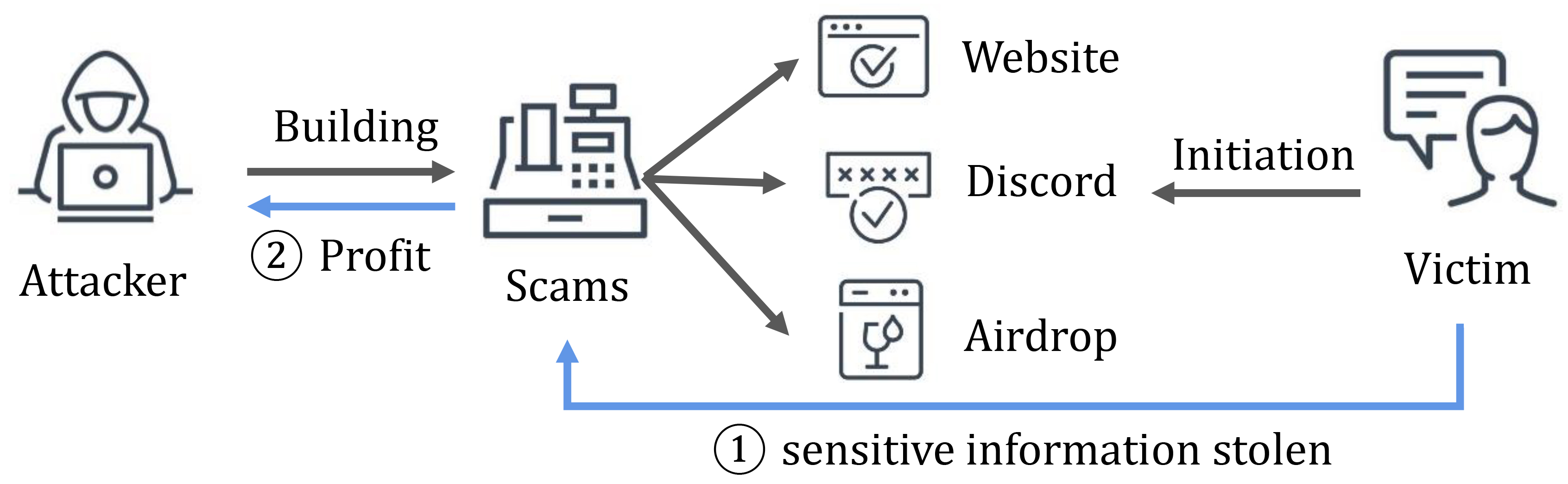}
\caption{A Motivating Example of Web3 scams on the Ethereum. The black line represents the activities under the chain, and the blue line indicates the behaviors on the chain.}
\label{fig:motivation}
\vspace{-1ex}
\end{figure}

Web3 scams refer to fraudulent or deceptive practices that exploit the network technologies, platforms, or services in Web3~\cite{liu2024fishing,wu2023towards,li2025uscsa}. Web3 scams can take various forms, such as Ponzi and Phishing. Web3 scams encompass phishing, however, there also exist discernible differences. As depicted in Fig. \ref{fig:motivation}, under the chain, traditional phishing is similar to web3 scams, which trick users into connecting their wallets to fake mediums such as websites and Discord~\cite{gao2025implementation,zhang2025attacks}. On the chain, traditional phishing accounts defraud users' funds~\cite{ding2025comprehensive}. However, the web3 scams pretend to offer regular services while allowing the attacker to gain access to the user's funds surreptitiously~\cite{liu2025empirical,sun2025data}. Subsequently, attackers could steal tokens without the user's involvement and obscure their traces~\cite{ase2024poster}. Thus, the primary disparity lies in the mimic behavior and track obscuration~\cite{yang2025multi,Wu2025security,zhang2025security,huang2025comparative,long2025fomo3d} (e.g., utilizing mixer contracts or multiple transfers).

In this paper, we focus on the multiple transfer behaviors performed by fraudsters to obscure traces. Thus, we analyze the external transactions related to the web3 scams. 
Previous works~\cite{kim2024drainclog,li2023siege,li2022ttagn,perozzi2014deepwalk,zhou2022behavior,yang2024stole}  have converted the Ethereum transactions into a multi-edge graph, which can intuitively represent the transaction patterns, helping identify malicious patterns. They attempt to learn node feature representations from the power-law distributed and temporal-attribute topology network graph, identifying malicious nodes or behaviors on the Ethereum network. However, limited research focuses on detecting malicious accounts or behaviors in web3 scams, and there are three significant challenges associated with graph learning methods for web3 scams.

\noindent$\bullet $ \textbf{Challenge 1 (C1):} \textit{\textbf{Large-scale Graph Construction:}} The transactions on Ethereum have grown to 2,762 million as of April 2025~\cite{etherscan2024}, which poses a significant challenge for deep learning methods that require processing the entire data. The majority of graph construction methods employ simple random sampling techniques, disregarding the attributes within the network~\cite{hu2023bert4eth,li2022ttagn,zhou2022behavior} (e.g., transaction sequence, top-k transaction graphs, random walk). 

\noindent$\bullet $ \textbf{Challenge 2 (C2):} \textit{\textbf{Temporal Series:}} The transaction within the network is inherent in time dependency, as evidenced by factors such as fund flow, and invocation. Learning the dynamic evolution in chronological order is crucial, given that interactions between malicious and victim nodes follow a specific sequence. However, most studies \cite{li2023siege,poursafaei2021sigtran,zhou2022behavior} only consider the network structures, neglecting the temporal series between the network structures. 

\noindent$\bullet $ \textbf{Challenge 3 (C3):} \textit{\textbf{Detection in the Web3 Domain:}} Although numerous detection tools for malicious accounts exist~\cite{hu2023bert4eth,li2023siege,li2022ttagn,poursafaei2021sigtran,zhou2022behavior}, only a few studies have addressed the scams in the web3 domain. Furthermore, there is a lack of research summarizing the current state of scams on web3, distinguishing them from other types of fraud nodes (e.g., phishing nodes), which hampers the development of malicious account detection in the specific domain of web3.

\noindent\textbf{Our Solutions.} To address these challenges, we introduce a temporal subgraph learning framework, called \textit{ScamSweeper}, to detect malicious accounts in web3 scams. ScamSweeper considers not only the graph structure but also the dynamic evolution feature between subgraphs. 

\noindent$\bullet $ \textbf{Solution to C1:} We take an optimized sample walking method to extract subgraphs from transactions, constructing subgraphs with temporal distribution. Each subnetwork is expanded from the walking sequences, where each node enriches a structure window-size edges, constructing directed graphs as a sampled dataset (\cref{sub:graph_construction}\footnote{In this paper, we use the symbol § to represent the meaning of Section, for example, § IV-C means the subsection C in Section IV.}). 

\noindent$\bullet $ \textbf{Solution to C2:} We partition each graph by temporal intervals to obtain multiple subgraphs and construct graphs with timestamps. We sort the subgraphs as a sequence to capture the dynamic evolution, thus gaining insights into the transaction time series (\cref{subsec:sequence_learning}). 

\noindent$\bullet $ \textbf{Solution to C3:} We observe the differences between malicious and normal nodes, particularly in the in- and out-degree distributions. Thus, we create directed graphs to facilitate the graph feature learning and capture the dynamic evolution of subgraphs constructed by the transactions between temporal intervals (\cref{subsec:graph_encoder}).

The main contributions of this paper are as follows:
\begin{itemize}
\item To the best of our knowledge, we are the \textit{first} to introduce the temporal subgraph sequence learning framework for identifying web3 scam accounts on Ethereum. We slice subgraphs in sequences via temporal intervals, analyzing the dynamic evolution of malicious nodes (\cref{sec::method}).

\item We enhance the sample walking method by integrating temporal attributes and structural information to extract subnetworks from extensive network data. The sampled subnetworks construct a directed graph, acquiring insights into the patterns of transaction graphs (\cref{sub:graph_construction}).

\item We collect and analyze a large-scale dataset on web3 scams, phishing, and normal accounts. We evaluate ScamSweeper through the dataset, achieving a weighted F1-score of 17.29\% higher than the state-of-the-art in web3 scam accounts and an F1-score of 17.5\% higher in phishing accounts (\cref{sec::evaluation}).

\item We open-source our codes and experiment data at \url{https://figshare.com/s/65d00a4c50c9d5188c06}.
\end{itemize}

\section{Related Work}
\label{sec::relatedwork}

In this section, we mainly introduce previous works on detecting malicious blockchain nodes. There have been various works on phishing scam account detection.

\subsection{Phishing Account Detection}
Several classification methods~\cite{zhang2023automatic} have been used in studies~\cite{chen2020phishing,farrugia2020detection,poursafaei2021sigtran,yuan2020detecting,chenliang2020phishing,hu2023bert4eth,li2023siege,li2022ttagn,liu2024fishing,yang2024stole,ratra2024graph,choi2024learning, kim2024drainclog} to improve the performance of phishing scam detection. 
Some works~\cite{choi2024learning, kim2024drainclog} leverage off-chain data to detect phishing scams, providing novel insights into domain information and social networks on social media.
Yang et al.~\cite{yang2024stole} analyze NFT phishing for the first time and summarize four phishing patterns and their corresponding behaviors.
Chen et al.~\cite{chen2020phishing} extracted 219 statistical features from the first- and second-order neighbors in the first 7 million blocks, utilizing the DElightGBM method to discover that the variance of the node's in-degree was the most influential factor in phishing node classification. 
S. Ratra et al.~\cite{ratra2024graph} and Farrugia et al.~\cite{farrugia2020detection} collect features of the graph network using GNN and statistics, then utilize the XGBoost and AdaBoost methods to classify phishing accounts. Moreover, the time difference in the network and the total transaction amount are ultimately recognized as the two most critical features~\cite{farrugia2020detection}.
Poursafaei et al.~\cite{poursafaei2021sigtran} utilized the Ri-walk approach, combining node structural features, neighbor features, and transaction network features as node features, and employed logistic regression for classification after integrating these features. 
Lastly, Yuan et al.~\cite{yuan2020detecting} employed Node2Vec to sample the network, learn node representations from the neighborhood, and then leverage the SVM algorithm for classification. 
Chen et al.~\cite{chenliang2020phishing} utilized statistical and structural features extracted from deep learning models to identify phishing nodes. They employed a graph convolutional network to learn the network structures and collected eight types of transaction features. 
Hu et al.~\cite{hu2023bert4eth} proposed a novel Transformer structure to detect malicious nodes. They embedded address information, location information, node type, degree information, transaction amount, and time information in the node-generated transaction sequences. The Transformer encoder was then used to generate the node embeddings for identification. 
Li et al.~\cite{li2023siege} proposed an incremental self-supervised learning method called SIEGE, which learned the spatial node features by observing the growth in the number of new neighboring nodes. They then focused on learning multiple graph features in temporal order and used both temporal and spatial features as node features. 
Li et al.~\cite{li2022ttagn} used temporal information by using GCN to understand the structural features of the graph and convert edge information into nodes. They trained on transactions and considered structural, statistical, and temporal features to identify phishing nodes. 
Different from the above studies, ScamSweeper samples the transaction network based on structural and temporal attributes, thereby reducing the computational resource consumption for large-scale transaction graphs.

\subsection{Web3 Scam Detection}
Regarding detecting scams on Web3, it can be divided into the smart contract and node levels. 

Smart contracts are a logical structure on the blockchain that can deploy logical operating code. web3 scams, such as honeypots~\cite{chen2020honeypot,hu2022scsguard}, Ponzi schemes~\cite{zheng2023securing}, and rug pull~\cite{huang2022graphlime}, can be implemented through smart contracts, leading to various detection methods. The n-gram model is leveraged to divide the bytecode into several groups, and then the LightGBM model~\cite{chen2020honeypot} is used to learn the features of the bytecode to detect honeypots. In addition, the attention mechanism~\cite{hu2022scsguard} is used for focusing the contributions of the groups and making a classification to detect honeypots and Ponzi schemes. Zheng et al.~\cite{zheng2023securing} employed a multi-view cascade ensemble model to detect Ponzi schemes. They combined the features of the contract bytecode, opcode, and deployer to achieve classification. Huang et al.~\cite{huang2022graphlime} detected rug pull~\cite{cernera2023token} by employing multi-dimensional features four days prior, including time series data, token transfer operations, and market trading features.

Node-level scam detection primarily focuses on identifying behavioral patterns in transactions or call operations. Kong et al.~\cite{kong2023defitainter} implemented taint analysis technology to analyze possible execution flows and observed stored state information in call data to detect vulnerabilities related to price manipulation. Varun et al.~\cite{varun2022mitigating} extracted features such as the gas price of transactions, the usage of gas tokens, and predicted gas prices from a large amount of transaction data. They used MLP to detect front-running vulnerabilities. Moreover, novel methods~\cite{hu2023sequence,la2023doge} are proposed for detecting pump-and-dump events. They processed the data by removing the moving standard deviation of rush orders, number, and volume of trades, etc.  
Hu et al.~\cite{hu2023sequence} combined statistical features (chain IDs and sequential transactions) to detect pump-and-dump events. The above methods concentrate on the static features of graph structure, which makes it difficult to detect accounts that mimic service providers. Different from these works, we not only optimize the sampling algorithm of the network graph, but also focus on the dynamic evolution of the transaction graph structure.


\section{Background}
\label{sec:background}
In this section, we provide the essential background on the malicious account detection, which aims to identify illicit accounts that behave maliciously in a transaction graph. 

\subsection{Sample Walking}
The sample walking method~\cite{grover2016node2vec, perozzi2014deepwalk} is a graph embedding technique that employs random walks to capture the local features of nodes in a graph. The method selects nodes randomly from the graph and generates node sequences by performing random walks. These node sequences are then used to learn the topological structure of the node from the neighbors. They analyze the neighborhood relationships between nodes in a graph by focusing on local features. Several graph embedding methods based on random walks are applied, such as Node2Vec~\cite{grover2016node2vec} and DeepWalk~\cite{perozzi2014deepwalk}. 

Random walk is a classical implementation of sample walking, which does not consider the attributes of edges in the network. Given a graph, $G=(V, E)$, which comprises a set of nodes represented by $V$ and a set of edges represented by $E$. It traverses a graph $G = (V, E)$ and collects a sequence of nodes $\{n_0,n_1,...,n_m\}$, where $n_x$ is the $x_{th}$ node, and $n_{x+1}$ is a randomly chosen neighbor of $n_x$ with some probability $p(n_{x+1}|n_x)$. Then, the node sequences are generated by performing random walks from each node. For each walk, the sample method randomly selected a neighbor of the current node and repeated this process until the walk reached a predefined length. After sample walking, node sequences are fed into the word embedding model, such as Word2Vec, to learn the node features. Moreover, DeepWalk assigns equal probabilities to the neighbors of each node and generates node feature vectors in the same dimension. Node2Vec leverages the alias sampling algorithm~\cite{grover2016node2vec} to sample edges, giving similar preference to all edges.

\subsection{Deep Graph Learning}
The deep graph learning method~\cite{buterez2025end} is a technique that aims to represent graphs in a lower-dimensional space to facilitate the analysis of graphs. A common approach in network representation is to utilize graphs as the basic units of representation, which requires two steps: graph extraction and graph encoding~\cite{yang2025staig}. Graph extraction selects relevant subgraphs from the input graph according to various criteria, such as k-core decomposition~\cite{kong2019k}, community detection~\cite{ma2023community}, etc. Graph encoding projects the high-dimensional features of the selected subgraphs to a low-dimensional vector space~\cite{kauffman2025infehr}. These different methods will capture the features of the graph and the relationships between subgraphs. Given a graph $G=(V, E, \mathbb{V}, \mathbb{E}, Y)$, it comprises a set of nodes represented by $V$ and a set of edges represented by $E$, the $\mathbb{V}, \mathbb{E}$ mean the node feature matrix and edge feature matrix, and the $Y=\{(v,y) | v\in V\}$ is the label set of the nodes, where $y$ is the label of node $v$. Graph neural networks~\cite{kipf2016semi,pan2023automated} can learn a function $F=\{f(g_v)\to y\}$, where $g_v$ is a subgraph derived from node $v$ in $G$, to get features from non-Euclidean data, such as graphs, networks, and point clouds. By jointly processing the attributes of the network, graph neural networks can capture both the structure and the attribute information of the graph~\cite{carvajal2025rnamigos2}. Subgraph-based statistics utilize the counts of various attributes in the subgraph as features to represent and analyze graphs, such as the motif sampling method~\cite{benson2016higher,wu2023know}. The resulting subgraph counts can capture the graph features.

\section{ScamSweeper}
\label{sec::method}

\begin{figure*}[!t]
\centering
\includegraphics[width=.86\linewidth]{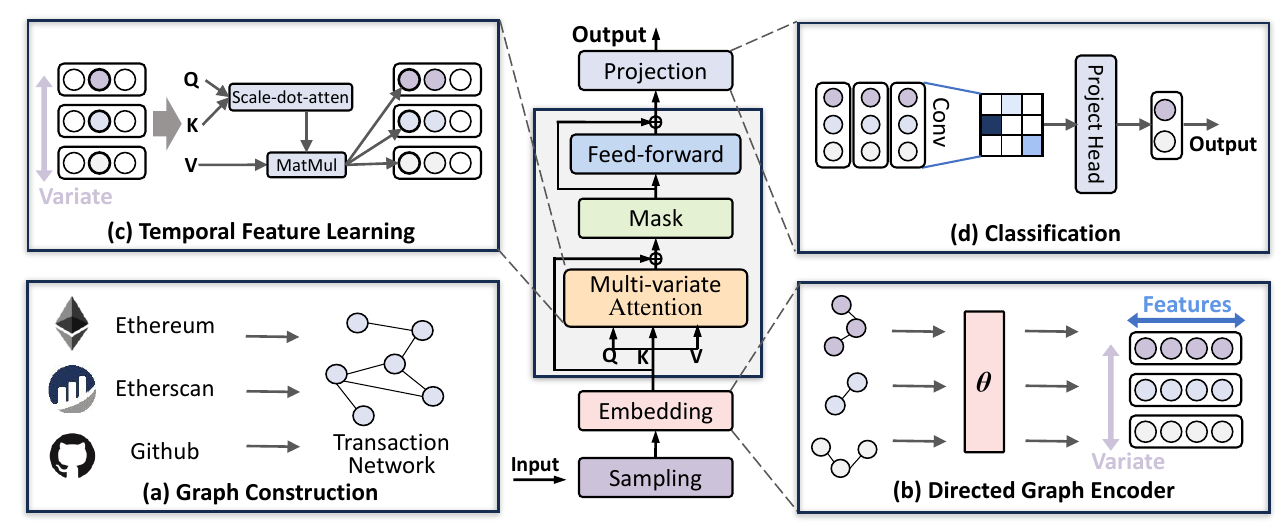}
\caption{The Main Example of ScamSweeper. It comprises four components that perform the following steps: (a) transactions are gathered from various public sources to construct a graph; (b) the features of subgraphs with temporal intervals are collected in the multi-directed graph; (c) self-attention is utilized to enhance the feature correlation of subgraphs across different temporal intervals; (d) deep neural network makes a classification with embeddings in local receptive fields.}
\label{fig:framework}
\end{figure*}

\subsection{Overview}

Our work comprises four steps, as illustrated in Fig. \ref{fig:framework}. Step (a) constructs a graph network by crawling transactions of accounts on Ethereum (\cref{sub:graph_construction}). The structure temporal random walk samples a graph network from transactions, building a dataset via time intervals as input (\cref{subsec:graph_encoder}). Step (b) extracts the graph features of the whole directed graph, assigning to each subgraph a directed graph encoder (\cref{subsec:graph_encoder}). Step (c) leverages a transposed Transformer to learn the time series features in the subgraph features (\cref{subsec:sequence_learning}). Step (d) employs a DNN method to make a classification for output (\cref{subsec:sequence_learning}). In the testing phase, ScamSweeper inputs the sampled network generated by the structure temporal random walk. Then, prediction results can be produced by steps (b), (c), and (d).

\subsection{Problem Description}
\subsubsection{\textbf{Definition}}
We model blockchain transactions as a graph, denoted as $G=(V,E)$. In the graph $G$, the account is expressed as a node\footnote{In this paper, we will use “account” and “node” interchangeably.}, and the transaction is expressed as an edge. The set of accounts forms the node set $V$, while the set of transactions constitutes the edge set $E$. The direction of the edges is consistent with the direction of the transactions.

\subsubsection{\textbf{Description}}
Since the graph learning method can intuitively learn the structure features of the node from the network graph, the sequence learning method can address the shortcomings of sequential temporal attributes in dealing with graph learning. Therefore, for a graph $G=(V, E)$ constructed by the blockchain network, we employ the sample walking method to extract the subgraph sequence $S = \{g_{v_0}^0,g_{v_0}^1,...,g_{v_0}^m\}$, where $g_{v_0}^i$ is the subgraph at the $i_{th}$ temporal interval, $v_0$ represents the start node, which is the source of the subgraph sequence, and $m$ is the length of the sequence. For the subgraph sequence $S$ of node $v_0$, we reacquire the node features matrix $\mathbb{V}$, edge features matrix $\mathbb{E}$, and the label $y$ of $v_0$, to construct a new graph $G_s=(V, E, \mathbb{V},\mathbb{E}, Y)$. Finally, we learn a function $f(g)\to y$ (i.e., \cref{subsec:graph_encoder} and \cref{subsec:sequence_learning}), where $g$ is a subgraph from the subgraph sequence $S$, mapping the subgraph to a low-dimensional label $y$ (i.e., malicious or benign).

\subsection{Graph Construction} 
\label{sub:graph_construction}
Before learning the features of transaction graphs, we need to obtain the transactions and construct the transactions as a network graph. We sample the original transaction network graph, generating the subgraph sequence from the account.

\subsubsection{\textbf{Transactions Retrieval}} \label{subsec:tx_collection}
Ethereum is a public trading web3 platform that enables users to deploy contracts executing complex transactions, such as decentralized applications (DApps), decentralized finance (DeFi), etc. Due to the multi-party property of these transactions, the number of transactions generated on web3 is higher than that of traditional Ethereum phishing transactions, which are only simple transfers of funds between accounts. For transactions, we obtain transaction records that contain timestamps, amounts, incoming addresses, and outgoing addresses, leveraging the BlockchainSpider tool~\cite{wu2023tracer}. Based on the large-scale transaction data collected, we construct a multi-directed graph. In this graph, the nodes represent Ethereum accounts, and the edges represent transactions between accounts. Moreover, edges exist between two nodes, each with a timestamp and transfer amount. A multi-directed graph is a type of graph structure, that allows multiple directed edges between the same pair of nodes.

\begin{figure}[ht]
\centering
\includegraphics[width=\linewidth]{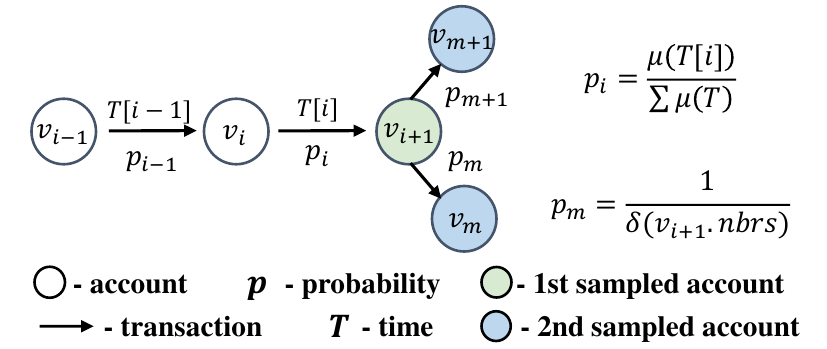}
\caption{An Example of Structure Temporal Random Walk. The green circle represents the first-step sampled node, and the blue circle indicates the second-step sampled nodes.}
\label{fig:temporalWalk}
\vspace{-3ex}
\end{figure}

\subsubsection{\textbf{Structure Temporal Random Walk}} \label{subsec:random_walk}
The transaction data on Ethereum has grown to 2,762 million until April 2025~\cite{etherscan2024}, which poses a significant challenge for deep graph learning methods that require processing the entire data. Therefore, to alleviate the requirements of large-scale transaction data on computing resources, we implement the Structure Temporal Random Walk (STRWalk) techniques to sample the network. It allows us to navigate the network and gather relevant neighbors and edges while streamlining the size and intricacy of the transaction network graph.

As Algorithm \ref{alg:structureTempoRalrandomWalk} shows, STRWalk takes the graph $G(V,E)$, start node $\upsilon$, and other parameters (i.e., structure window size $w$, temporal interval $\tau$, walk length $\xi$) as input. When the number of walks satisfies the walk length $\xi$ in Line 12, it will return the sampled graph as output. Through inputting different start nodes, such as phishing or normal accounts, we can obtain corresponding sampled graphs and use them to train and evaluate ScamSweeper in \cref{sec::evaluation}.

STRWalk can be divided into two steps. Firstly, STRWalk utilizes the temporal attribute, which is an indispensable element of transactions. To account for the temporal dimension of many contemporary problems~\cite{wu2023understanding,li2022ttagn,hu2023bert4eth}, we incorporate the temporal attributes into the node sequence sampled from the network. During Line 14-23, given the current node $v_i$, it gets the neighbors $\{v_i^1,v_i^2,...,v_i^n\}$ of $v_i$, it computes the set of temporal attributes $T=\{t^1,t^2, ...,t^n\}$ for all the edges connected neighbors of $v_i$, where $t^j \in T$ means the timestamp value of edge between $v_i$ and $v_i^j$, $j \in [1,n]$. During each walk, we calculate the probability $P_i$ as shown in Eq.\ref{eqa:temporal_walk_p}, which subtracts the minimum value in $T$ from each element.

\begin{equation}
\begin{aligned}
   P_i &= \frac{\mu(T[i])}{\sum_{j=1}^n \mu(T[j])},\\
   \mu(T[\ell]) &= T[\ell] - \text{Min}(T) + 1
\label{eqa:temporal_walk_p}
\end{aligned}
\end{equation}
where $e_i$ means the edge between $v_i$ and $v_{i+1}$, $T[i]$ indicates the timestamp when the edge $e_i$ happened, $\mu$ represents the difference operation from the shortest time of edges in the network. Through the Eq.\ref{eqa:temporal_walk_p}, edges closer to the start node will get higher sampling weights. The purpose of the Eq.\ref{eqa:temporal_walk_p} is edges closer in time to the start node should be given more weight, since the closer edge is more effective at representing the start node~\cite{wu2023understanding}. 
Then, we employ the typical alias sample algorithm~\cite{perozzi2014deepwalk} to randomly select a neighbor of $n_i$ according to the probability distribution $P$. Under the first step of STRWalk, it weights the temporal property of the network.

\begin{algorithm}[ht]
\caption{StructureTemporalRandomWalk($G,w,\tau,\upsilon,t$)}\label{alg:structureTempoRalrandomWalk}
\small
\SetAlgoNoLine
\normalem
  \SetKwData{Left}{left}\SetKwData{This}{this}
  \SetKwProg{Fn}{Function}{:}{}
  \SetKwFunction{getStruct}{Structure\_Sample} 
  \SetKwInOut{Input}{Input}\SetKwInOut{Output}{Output}

	\Input{graph $G=(V,E)$; structure window size $w$;\\ temporal interval $\tau$; start node $\upsilon$; walk length $\xi$;}
	\Output{sampled graph $G_s$}
	\BlankLine
    \Fn{\getStruct{$\upsilon,G$}}{ 
        $V,E= []$   \;        
        $\sigma =$ GetMinTimestamp($\upsilon, G$)\;
        $edges$ = SearchEdges($\upsilon, \sigma, \sigma+\tau$); \Comment{get the edges during $(\sigma, \sigma+\tau)$}\; 
        \For{$i \in [1,w]$ }{ 
            $E \leftarrow $ Alias\_Sample($1/$Num$(edges)$); \Comment{sample $w$ nodes}\;
        }
        $V \leftarrow $ next\_node($E$)\;
        \KwRet $(V,E)$\;
    }
    
    walk = [$\upsilon$]        \Comment{Initiate the walk sequence}\;

    $G_s \leftarrow$\getStruct{$\upsilon,G$}\;
    
    \For{$i \in [1,\xi)$}{
        $v_i$ = walk[-1];  \Comment{get the current node $v_i$}\;
        Get the neighbors $v_i.nbrs$ of the current node $v_i$, and the loop breaks if $v_i.nbrs$ equals to $0$\;
        Get the edges $\epsilon$ connected $v_i$ and its neighbors $v_i.nbrs$\;
        $\sigma =$ GetMinTimestamp($\upsilon, G$)\;
        Sort the edges $\epsilon$ by time in ascending order to get $\lambda$\;
        $\varepsilon$ = Alias\_Sample($\lambda/\sum \lambda$)\;
        $v_{i+1}$ = next\_node($\varepsilon$);  \Comment{get another node in the edge}\;
        walk.append($v_{i+1}$)\;
        $G_s \leftarrow$\getStruct{$v_{i+1},G$}\;
    }
    \KwRet $G_s$ 
    
\end{algorithm} 

As for the second step of STRWalk in Line 27, which will be equivalent to the function in lines 1-10, enriching the structural information of the nodes. The purpose of this step is that although the temporal attributes are extracted through the first step, the structure information is also important for graph learning. Therefore, we apply an expansion strategy to the walking sequence, extending the structure information. To augment the structural information of the nodes, Fig. \ref{fig:temporalWalk} shows that we supplement other nodes. Suppose that the current node $v_i$ has finished the first step of STRWalk, and starts the second step. It selects the related nodes at the same temporal interval $\tau$. The probability distribution can be calculated as Eq.\ref{eqa:struc_temporal_walk_p}, and then the alias sample algorithm will be used for selection.

\begin{equation}
   P_m = \frac{1}{\delta(v_i.nbrs)} 
\label{eqa:struc_temporal_walk_p}
\end{equation}
where $\delta(v)$ represents the number of nodes that are in the same interval with node $v$.  

Moreover, we set the time unit to \textit{day} to avoid collecting useless information with short intervals and prevent noise with long intervals~\cite{wu2023understanding}. Suppose that the interval is $k$ days, the position of the temporal interval $\tau$ can be computed as following Eq.\ref{eqa:time_interval_id}, where 86,400 represents one day, as the timestamp is measured in seconds.

\begin{equation}
\begin{aligned}
    \tau =\frac{t - t_{first}}{86\text{,}400*k} 
\label{eqa:time_interval_id}
\end{aligned}
\end{equation}

Then, we repeat the first and second steps until the sampled walk sequence reaches a fixed length or the nodes have no neighbors. Finally, the node sequence $S_w=\{v_0,v_1,...,v_m\}$ is selected by the first step, and the second step obtains the subgraph sequence $\{g_{v_0}^0,g_{v_0}^1,...,g_{v_0}^m\}$, where $g_n^i$ from the $v_i \in S_w$ and the $v_0$ is the start node. Therefore, the expansion strategy integrates the structural features into the node sequence and forms the subgraph sequence as the sampled graph $G_s$.

\subsection{Directed Graph Encoder}
\label{subsec:graph_encoder}

Given the sampled graph $G_s=(V,E)$ with the start node $v_0$ contains $m$ subgraphs $\{g_{v_0}^0, g_{v_0}^1, ..., g_{v_0}^m\}$ as sequence. We utilize the directed graph encoder shown in Fig. \ref{fig:encoder} to learn the structure feature of each subgraph in $G_s$.

\subsubsection{\textbf{Neighborhood Feature Construction}} \label{subsec:directed_graph_construction}
Let the subgraph $G_g$ as input graph with $\alpha$ nodes $\mathcal{V}=\{v_1,...,v_\alpha\}$ and $\beta$ edges $\mathcal{E}=\{e_1,...,e_\beta\}\subseteq \mathcal{V}^2$, associated with node features $X^\mathcal{V} \in \mathbb{R}^{\alpha \times C}$ and edge features $X^\mathcal{E} \in \mathbb{R}^{\beta \times C}$. Then, we construct the edge feature matrix $\mathbb{E}$ and the node feature matrix $\mathbb{V}$, converting the node sequence network $G=(V, E)$ to $G=(V, E,\mathbb{V},\mathbb{E}, Y)$ for graph learning.

\begin{figure*}[ht]
\centering
\includegraphics[width=\linewidth]{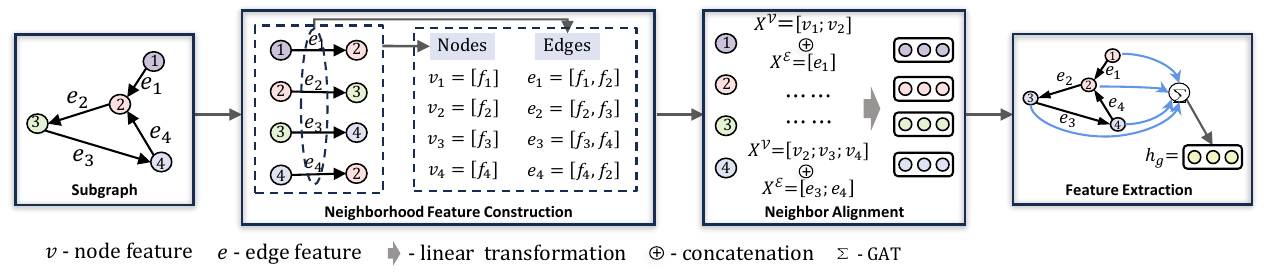}
\caption{An Example of Directed Graph Encoder in a Single Temporal Interval.}
\label{fig:encoder}
\end{figure*}

Given a node walk sequence in the network, we consider that multiple edges can be between any pair of nodes, and each edge connects two nodes. For all transactions $E=\{X_f \to X_t\}$, $X_f$ and $X_t$ are the sets of source and target node features, respectively, which can be defined as $X_f = [X_f^1, X_f^2, ..., X_f^n]$, $X_t = [X_t^1, X_t^2, ..., X_t^n]$, where $X_f^i$ represents the $i_{th}$ incoming node feature, $X_t^i$ represents the $i_{th}$ outgoing node feature, and $n$ is the edge amount. The nodes and edges representation set in the graph can then be represented as following Eq.\ref{eqa:node_edge_representations}, where $e^i$ means the $i_{th}$ transaction.

\begin{equation}
\begin{aligned}
    V &= \{X_f; X_t| (X_f^1;X_t^1, X_f^2;X_t^2, ..., X_f^n;X_t^n)\}\\
    E &= \{X_f\to X_t|(e_1, e_2, ..., e_n)\}
\label{eqa:node_edge_representations}
\end{aligned}
\end{equation}

To construct the edge feature matrix $\mathbb{E}$, we use a matrix $W \in \mathbb{R}^{1 \times n}$ to represent $\mathbb{E} = WE$. Moreover, the node feature matrix can be repressed by $\mathbb{V}= [WX_f, WX_t]^T$. Utilizing the directed graph encoder $\theta$ in Fig. \ref{fig:framework}(b), we can learn the features of all the nodes and the corresponding edges in the latent space from the whole graph. For a node $v$, the derived network is represented by $G=\{g_{v}^0,g_{v}^1,$ $...,g_{v}^m\}$, where $g_{v}^i$ is the subgraph at $i_{th}$ interval, and $m$ is the number of intervals that a node network have been survived. The learning process of the subgraph is illustrated in Fig. \ref{fig:encoder}.

\subsubsection{\textbf{Neighbor Alignment}}
\label{subsec:neighbor_alignment}
As the subgraph is centered on a certain start node, we need to examine the features of its neighbors and the edges, to identify the patterns and features of the graph. In the neighborhood feature construction process, we first obtain all the node and edge representations in the whole graph. Then, for each subgraph, we get all the node features $X^\mathcal{V}$ and the directed edge features $X^\mathcal{E}$ sequentially concatenated by the two nodes. For example, as for the node $f_1$ in the subgraph $G_g$ in Fig. \ref{fig:encoder}, the $X^\mathcal{V}$ is the concatenation of related node features $[v_1;v_2]$, the $X^\mathcal{E}$ is the concatenation of related edge features $[e_1=[f_1;f_2]]$.

In each subgraph, the nodes contain different numbers of neighbors. Therefore, the dimensions of neighborhood features $X^\mathcal{V}$ and directed edge features $X^\mathcal{E}$ are different. To align these differences, we normalize the attribute features of the neighbor nodes, and Eq. \ref{eqa:neighborhood_features} aligns the neighborhood features of the associated target node.

\begin{equation}
\begin{aligned}
    \hat{v} = \text{LeakyRelu}(\Theta_v \cdot [X^\mathcal{V} \vert X^\mathcal{E}] + b_0),
\label{eqa:neighborhood_features}
\end{aligned}
\end{equation}
 
\noindent where $\Theta_v$ and $b_0$ are a linear transformation layer and the bias respectively, which are employed to align the dimensions of the neighbor features. For nodes with various numbers of neighbors, the zero padding is used to unify their dimensions.

\subsubsection{\textbf{Feature Extraction}}
\label{subsec:feature_extraction}
The subgraph is characterized by its node-centric structure, in which each constituent node in the neighborhood possesses a different magnitude of significance to the central node. 
To effectively extract the feature of the central node from the perspective of the subgraph, we need to comprehend the different significance of the neighborhood with the source node. Hence, to quantify the neighborhood with different contributions, we employ the graph attention network (GAT)~\cite{velivckovic2017graph} to delineate the behavior patterns. It facilitates calculating the importance degree among nodes and learning graph-pattern features, thereby learning the structure feature in each subgraph.

Specifically, for the central node $v_a$ in the subgraph $g_{v_a}$, the GAT network first obtains the attention scores of the neighbor $v_b$ for the $v_a$ in the subgraph $g_{v_a}$ as following Eq.\ref{eqa:importance_obtain},

\begin{equation}
\begin{aligned}
    e_{ij} = \text{LeakyRelu}(\Theta_n \cdot [h_i \vert h_j]),
\label{eqa:importance_obtain}
\end{aligned}
\end{equation}

\noindent where $h_i$ and $h_j$ are the hidden features of $v_a$ and $v_b$, and a linear transformation $\Theta_n$ and an activation function $LeakyRelu$ are performed to compute the attention scores. The initial value of $h_i$ is the feature of the subgraph's central node $v_a$. Then, to compare with all the neighbors, we employ a softmax layer to make a normalization with the scores as following Eq.\ref{eqa:softmax},

\begin{equation}
\begin{aligned}
    \alpha_{ij} = \frac{\text{exp}(e_{ij})} {\sum_{x\in N(i)} \text{exp}(e_{ix}) } ,
\label{eqa:softmax}
\end{aligned}
\end{equation}

\noindent where $N(i)$ is the neighbors of $v_a$. After obtaining the importance distribution, we incorporate it into the node features to update the feature of the subgraph as following Eq.\ref{eqa:result},

\begin{equation}
\begin{aligned}
    h_{g} = \text{Elu}(\alpha_{ij} \cdot \Theta \cdot h_i + \sum_{x\in N(i)} \alpha_{ix} \cdot \Theta \cdot h_x ) ,
\label{eqa:result}
\end{aligned}
\end{equation}

\noindent where $h_g$ represents the feature of the subgraph $g_v$, $h_i$ is the node feature of subgraph's central node $v_a$, $Elu$ is an activation function~\cite{hendrycks2016gaussian}, $\Theta$ means a linear transformation layer.  

By extracting the feature $h_g$ of each subgraph $G_g$ in all the temporal intervals sequentially, we can obtain the temporal subgraph feature sequence $\Phi=(h_g^0,h_g^1,...,h_g^m)\in \mathbb{R}^{m\times D}$ for \cref{subsec:sequence_learning}, where $D$ means the hidden size, $m$ is the variable temporal length.

\subsection{Temporal Subgraph Sequence Learning} \label{subsec:sequence_learning}
\begin{figure}[!t]
\centering
\includegraphics[width=\linewidth]{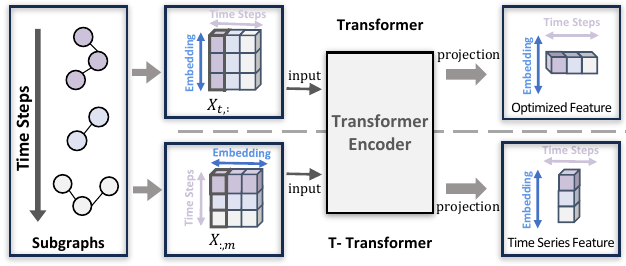}
\caption{A Difference Example of the Transposed and the Traditional Transformer. The upper part of the dotted line is the transposed Transformer, while the lower part represents the traditional Transformer.}
\label{fig:t-Transformer}
\end{figure}

To capture the dynamic evolution in the temporal subgraph feature sequence in \cref{subsec:graph_encoder}, we employ sequence learning techniques as shown in Fig. \ref{fig:t-Transformer}. 

The traditional Transformer~\cite{vaswani2017attention} learns the features of different time steps in the embedding dimension. However, to learn the dynamic evolution of the embedding over the whole time steps, we employ transposed Transformer layers for learning, which has been proven to be effective at \cref{subsec:RQ2}. For example, after extracting the features in \cref{subsec:graph_encoder}, we divide the entire temporal graph into several intervals (see Eq. \ref{eqa:time_interval_id}), resulting in a variate time series $G=(g_{n}^0,g_{n}^1,...,g_{n}^m)$ and a variate time feature $\Phi=(h_g^0,h_g^1,...,h_g^m)\in \mathbb{R}^{m\times D}$, where $D$ means the hidden dimension, $m$ is the variable temporal length. The dimension of each subgraph feature is fixed, assuming that $X_{t,:}$ is the variable positions at the same temporal interval, and $X_{:,m}$ is the entire time series at the same position. Fig. \ref{fig:t-Transformer} illustrates that the Transformer~\cite{vaswani2017attention} learns feature sequences at the same time from $X_{t,:}$ simultaneously. To learn the dynamic evolution of features, we focus on $X_{:,m}$, which captures the temporal evolution at each position of embedding. The transposed Transformer structure transposes the feature matrix to learn the same feature at different time steps. As shown in Fig. \ref{fig:framework}, we utilize the transposed layers to extract the feature representation $H=\{h_0,h_1,...,h_{m-1}\}\in \mathbb{R}^{m \times d}$ from $X_{:,m}$, where $h_i \in \mathbb{R}^{d}$ captures the temporal variation of the entire temporal sequence. 

First, we feed the time series into linear layers $\Theta_Q$, $\Theta_K$, and $\Theta_V \in \mathbb{R}^{d \times d}$, generating queries $Q$, keys $K$, and values $V\in \mathbb{R}^{m\times d}$. Then, the self-attention layer computes the importance scores of multiple tokens at the same time step, which can weigh the next hidden state in $V$, as following Eq.\ref{eqa:t-Transfomer},

\begin{equation}
\begin{aligned}
    h_s &= \text{softmax}(\frac{Q \cdot K^T}{\sqrt{d}}) \cdot V,
\label{eqa:t-Transfomer}
\end{aligned}
\end{equation}

\noindent where $h_s$ is the output, which is the temporally weighted vector at spatial locations across varying temporal dimensions. 

Furthermore, a mask mechanism is employed. The mask approach uses a vast negative or minor positive number to fill specific positions, resulting in a zero value at some positions during weight calculation. The mask approach can prevent the self-attention calculation from capturing future information, and disregard the padding position. Moreover, to enrich the feature, we implement the feed-forward layer, which is practically a two-layer fully connected network that can be calculated by Eq.\ref{eqa:feed-forward}:

\begin{equation}
\begin{aligned}
    h = (\sigma(h_s \cdot W_1+b_1))\cdot W_2 + b_2,
\label{eqa:feed-forward}
\end{aligned}
\end{equation}

\noindent where $W_1, W_2 \in \mathbb{R}^{d\times d}$ and $b_1, b_2 \in \mathbb{R}^{d\times 1}$ are the trainable parameters, $\sigma$ means the Sigmoid activation function. $h$ is the result of the feed-forward layer.

Then, we employ a deep neural network to discern temporal sequence relationships across different locations. For example, the convolutional neural network (CNN) and pooling layers~\cite{zeng2023financial} learn features at different locations for classification, which is predicated on the widely acknowledged ability of classification tasks. Lastly, the Projection head~\cite{vaswani2017attention} is used to make the classification.

Furthermore, during the training process, we aimed to minimize the objective function: $O(\theta) = L(\theta) + \omega(\theta)$, where $L(\cdot)$ represents the cross-entropy loss function and $\omega(\cdot)$ denotes the regularization term to prevent overfitting.

\section{Experimental Evaluation}
\label{sec::evaluation}
In this section, we conduct experiments to evaluate ScamSweeper and address the following research questions (RQs):

\noindent RQ1: \textbf{What distinguishes normal and malicious accounts?}
What is the difference between malicious accounts and normal accounts, and why is ScamSweeper effective?\\ 
\noindent RQ2: \textbf{Is ScamSweeper effective for detecting Web3 scams?}
Will ScamSweeper work better than other methods?\\ 
\noindent RQ3: \textbf{Is each component in ScamSweeper effective?} 
Is STRWalk robust? Which component (graph learning or sequence learning) makes a greater contribution to ScamSweeper?\\
\noindent RQ4: \textbf{Is ScamSweeper effective for detecting phishing accounts?} 
Will ScamSweeper have better detection capabilities than existing phishing account detection methods?\\
\vspace{-2ex}

\subsection{Experimental Setup}
\label{subsubsec:setup}
\subsubsection{\textbf{Datasets}}
\label{subsubsec:dataset}
We collect a large-scale transaction dataset of phishing and web3 scam accounts for evaluation, two types of malicious accounts on the Ethereum blockchain. Leveraging the BlockchainSpider tool~\cite{wu2023tracer} with the depth-first algorithm, we scrape the first 18 million blocks, covering the period from July 2015 to November 2023. During this period, the transactions of 4,905 phishing accounts labeled by Etherscan are crawled in this dataset. We also supplement the dataset with the 3,125 transaction networks from the web3 scam Database~\cite{scamsniffer2024database}, which exhibits scam behavior while providing web3 services. All the data are labeled and audited by the researchers, before collecting them through an official API from the Etherscan, thus ensuring the soundness of the dataset.

We randomly divide the dataset at \cref{subsubsec:setup} into 70\%, 20\%, and 10\% for training, validation, and test set, respectively.

\subsubsection{\textbf{Baseline}}
We compare ScamSweeper with several studies~\cite{yuan2020detecting,choi2024learning,chenliang2020phishing,zhou2022behavior,li2023siege,wang2023phishing,hu2023bert4eth}. They can divided into three types, including the random walk-based method (i.e., Yuan et al.~\cite{yuan2020detecting} and Choi et al.~\cite{choi2024learning} utilize the Node2vec and DeepWalk, respectively), the graph learning method (i.e., the GCN~\cite{kipf2016semi}, GAT~\cite{velivckovic2017graph}, GraphSAGE~\cite{hamilton2017inductive} which are embedded into Chen et al.~\cite{chenliang2020phishing}, Zhou et al.~\cite{zhou2022behavior}, and Li et al.~\cite{li2023siege}, respectively), and the sequence learning methods (i.e., PDTGA~\cite{wang2023phishing} and BERT4ETH~\cite{hu2023bert4eth} leverages the Transformer structure).

Regarding the random walk-based methods, Node2Vec~\cite{grover2016node2vec} is a method that uses the random walk technique to learn the feature representation of accounts, which are represented as nodes in the graph or network. Yuan et al.~\cite{yuan2020detecting} leverage Node2Vec to learn the domain representation, where the node network is generated by random walk and skip-gram. Similar to Node2Vec, Choi et al.~\cite{choi2024learning} adopt DeepWalk~\cite{perozzi2014deepwalk}, which utilizes a random walk to obtain graph networks, then generates features of nodes by the Word2Vec method.

Regarding the graph learning method, Chen et al.~\cite{chenliang2020phishing} evaluate the malicious account identification task with the GCN method, which refers to a kind of neural network that performs convolution operations on a graph. It gives weight to the features of nearby nodes as node features. Different from GCN, GAT~\cite{velivckovic2017graph} is a graph neural network that introduces an attention mechanism. Zhou et al.~\cite{zhou2022behavior} utilizes GAT to weigh the features of neighboring nodes, so that the model concentrates more on the appropriate neighbors. In contrast to GCN and GAT, GraphSAGE~\cite{hamilton2017inductive} is optimized to generate embeddings for previously unseen data, allowing it to handle dynamic graphs.

Regarding the sequence learning method, the PDTGA~\cite{wang2023phishing} leverages the Transformer~\cite{vaswani2017attention} structure, which is a prominent sequence learning model with the attention mechanism, capturing the dependencies of lengthy sequences. The encoder receives the input sequence, while the decoder generates the output sequence. 
The tool named BERT4ETH~\cite{hu2023bert4eth} is a Transformer-based neural network architecture that has been designed and fine-tuned explicitly on Ethereum blockchain datasets. Empirical evidence has established its efficacy in identifying and preventing the operation of phishing accounts.

\subsubsection{\textbf{Evaluation Metrics}}
In this paper, we mainly leverage the F1-score to evaluate our model's ability to detect malicious nodes on Ethereum. To gain a more comprehensive analysis of the model, we also employ accuracy, precision, and recall for evaluation. 
The accuracy metric indicates the percentage of correct predictions made by the model, encompassing both positive and negative predictions.
The recall metric reveals the likelihood of actual web3 scams or phishing nodes in the test samples being accurately identified by the model.
The precision metric demonstrates the ratio of true positives in the test samples that are correctly identified as positive by the model.
The F1-score is a comprehensive performance metric that considers both precision and recall. It reflects the disparity between the model's classification outcomes and the actual values and is the harmonic mean of precision and recall. Higher F1-scores indicate superior classification performance.

\subsubsection{\textbf{Environments}}
We perform the analysis and experiments on an Ubuntu server 22.04, equipped with an NVIDIA GeForce GTX 4070TI GPU and an Intel(R) Core(TM) i9-13900KF CPU. The server boasts 128GB of RAM and 1TB of SSD storage. To learn the features of transactions, we collected transactions from labeled malicious and normal accounts during the data collection phase. To collect second-order transactions, we utilize the breadth-first algorithm and set the query timeout to 180 seconds. In the STRWalk method, we set the structure window to \{5,10,15\}. For optimization, we use the Adam and set the weight decay rate to $ 5 \times 10^ {- 4} $.

\subsection{RQ1: Difference among Normal, Phishing, Web3 Scam Accounts} \label{subsec:RQ1}
We conduct an analysis of the distribution phenomena among normal, phishing, and web3 scam accounts, fostering a deeper understanding and examination of the distinctions between malicious and normal accounts.

We collect normal, phishing, and web3 scam nodes on Ethereum shown at \cref{subsubsec:setup}. The collection method of normal accounts follows Wu et al.~\cite{wu2023understanding}, including the classes of exchange, mining, ICO wallet, and gambling, which is marked in xlabelcloud~\cite{xblock2024}. After the collection, we construct a multi-directional graph by treating the accounts as nodes and the transaction vectors between the accounts as directed edges. The attributes of the multi-directional graph show in Table \ref{tab:dataset_analysis}.

\begin{table}[ht]
\small
\centering
  \caption{Statistical Information in the Multi-directional Graph. \#Nodes means the total number of nodes, \#L is the number of source nodes, \#Edge indicates the number of transactions, and \#SD Degree is the average variance value.}
  \label{tab:dataset_analysis}
  \scalebox{0.92}{
  \begin{tabular}{lllll} 
    \toprule
    \textbf{Datasets} & \textbf{\#Nodes} & \textbf{\#L} & \textbf{\#Edges} & \textbf{\#SD Degree}\\
    \midrule
    Normal         &  12,042,066  &  636  &  142,750,370   &  3555.35 \\
    Phishing       &  10,159,847 &  4,905  & 62,011,219    &  1285.25 \\
    Web3 Scams     &  8,736,430  &  3,125  &  64,265,586   &  541.10 \\
  \bottomrule
\end{tabular}
}
\end{table}

\vspace{-4ex}
\begin{figure}[hbp]
\centering
\subfloat[\small{Normal}]{\includegraphics[width=.33\columnwidth]{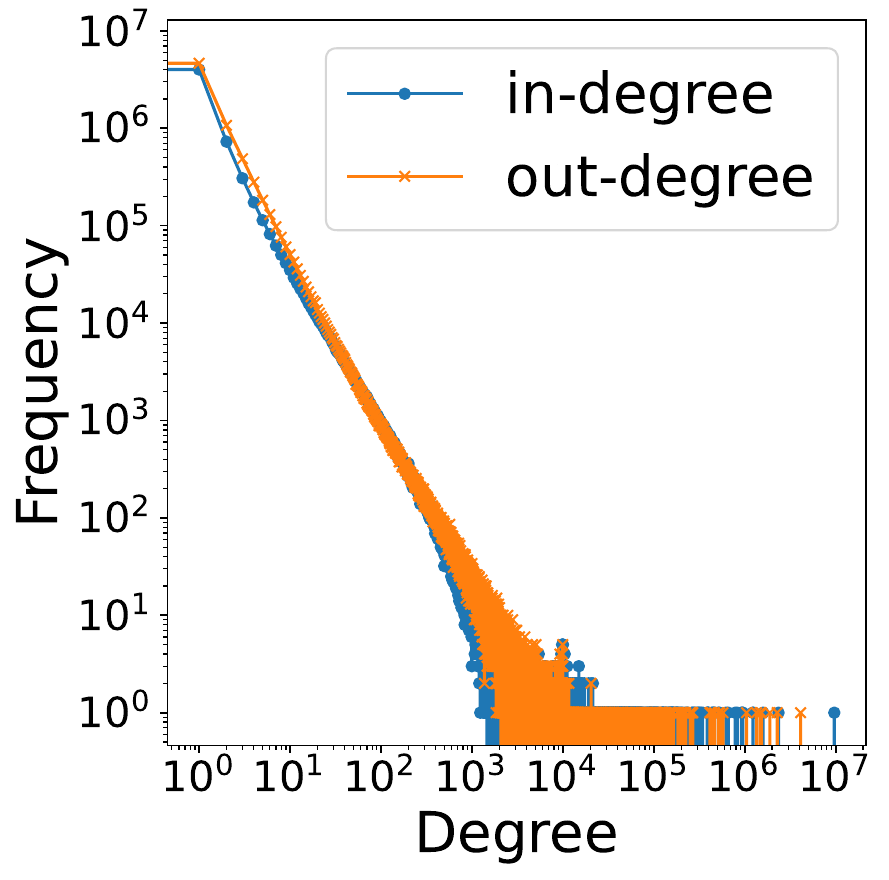}}
\subfloat[\small{Phishing}]{\includegraphics[width=.33\columnwidth]{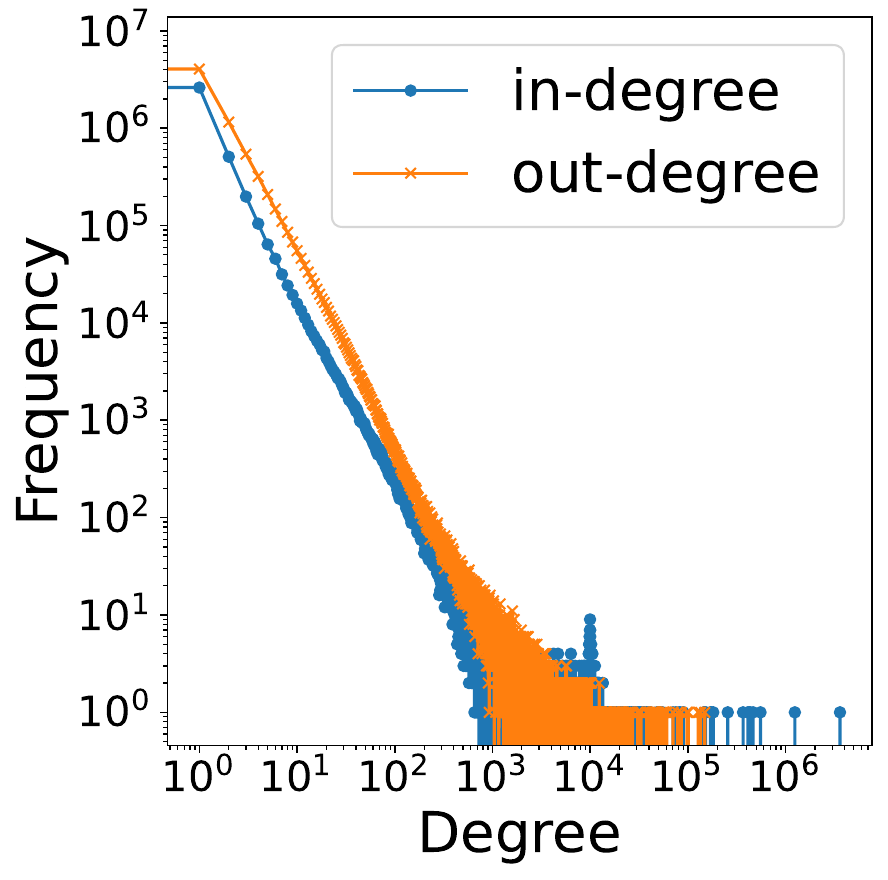}}
\subfloat[\small{Web3 Scams}]{\includegraphics[width=.33\columnwidth]{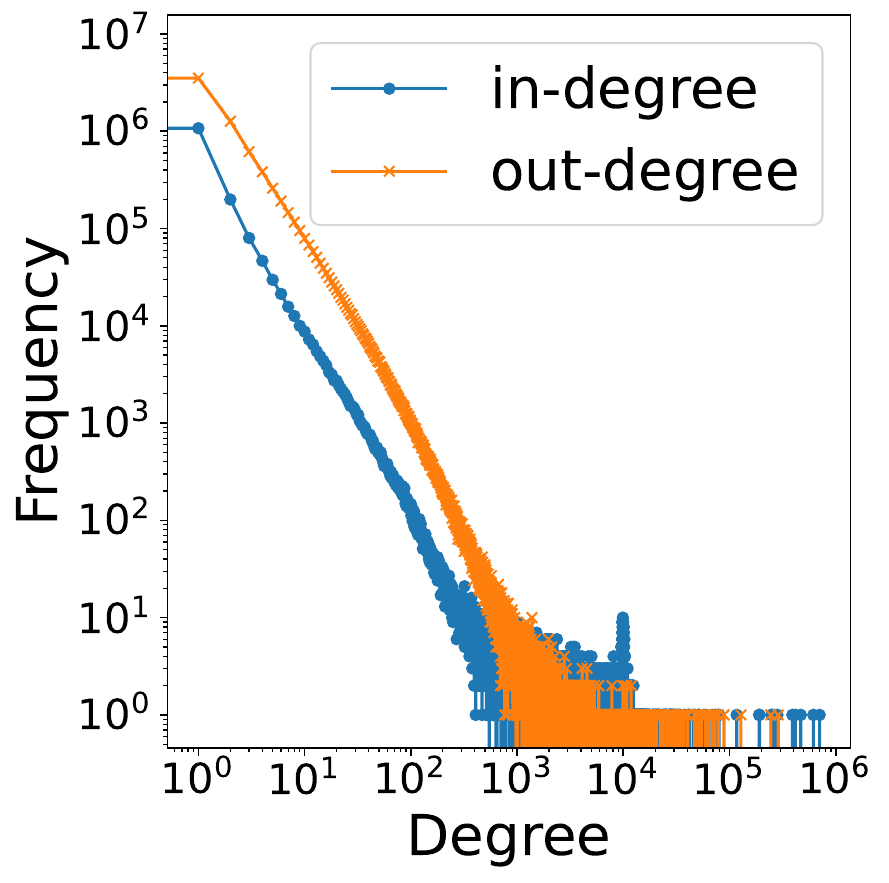}}\\
\vspace{-1ex}
\caption{The Frequency Distribution of in- and out-Degrees for Nodes Classified as Normal, Phishing, and Web3 Scams. The blue dot represents the in-degree of the nodes, and the orange dot represents the out-degree of the nodes.}
\label{fig:Frequency_Distribution}
\end{figure}

The dataset analysis provided in Table \ref{tab:dataset_analysis} reveals that the network of normal nodes boasts a greater average number of edges and nodes when compared to other networks. However, the network of normal nodes also exhibits the largest standard deviation of degrees, which is indicative of a more dispersed node distribution. On the other hand, web3 scams showcase the smallest standard deviation of degrees, suggesting a more clustered node distribution where a limited number of nodes generate most transactions.

To visualize and compare the different nodes, we draw a degree-frequency graph in Fig. \ref{fig:Frequency_Distribution} to analyze the degrees of the nodes. The degree distribution of all categorized nodes, as depicted in Fig. \ref{fig:Frequency_Distribution}, adheres to a power-law distribution. The long-tail phenomenon posits that most of the revenue is concentrated among these few critical nodes. Upon examination of the disparity between in-degrees and out-degrees, it becomes apparent that the entry and exit degrees of standard nodes are comparably similar. However, there is a discernible difference in the access degrees between phishing nodes and those associated with web3 scams, with the latter exhibiting a more pronounced effect. Additionally, the frequency of consumption (out-degree) transactions significantly surpasses that of profit (in-degree) transactions. The pattern implies that these nodes primarily generate income via extensive outward propagation. Furthermore, the overall number of incoming connections to nodes in the phishing network is lower than in web3 scams. Conversely, the phishing network has a higher total out-degree of nodes compared to web3 scams. 

\begin{figure*}[ht]
\centering
\subfloat[\small{Accuracy}]{\includegraphics[width=.5\columnwidth]{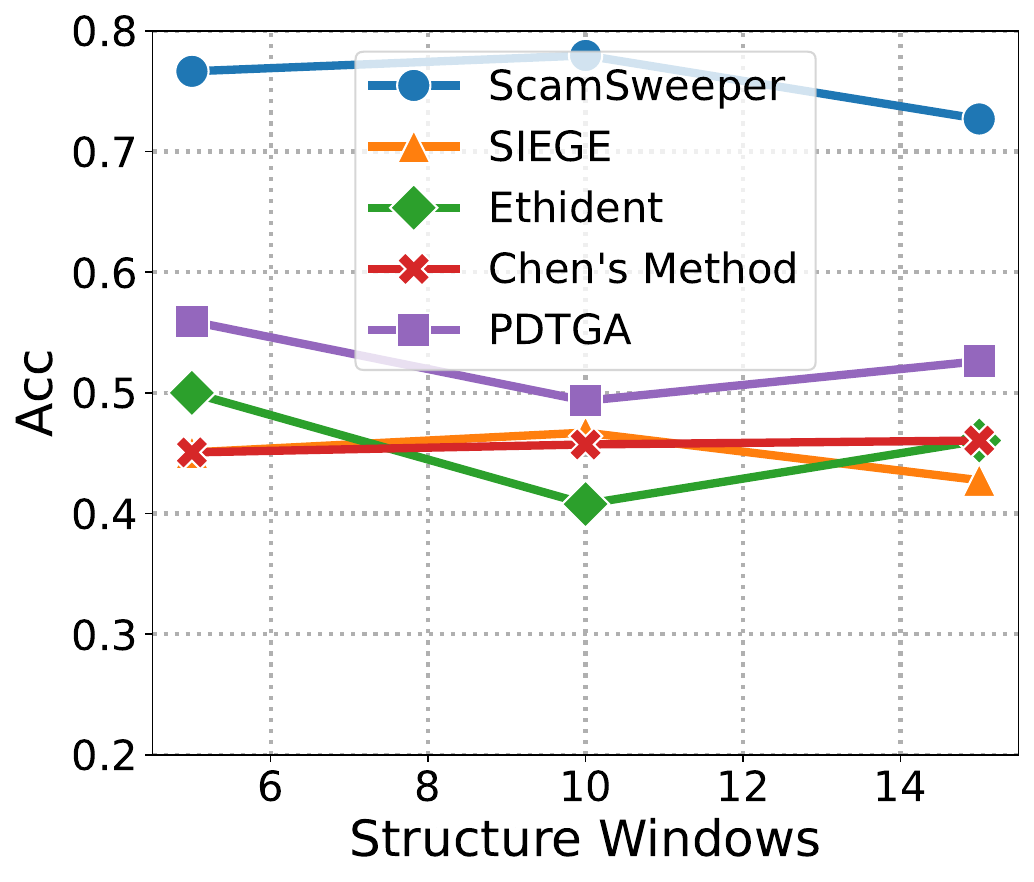}}\hspace{0.5pt}
\subfloat[\small{F1-score}]{\includegraphics[width=.5\columnwidth]{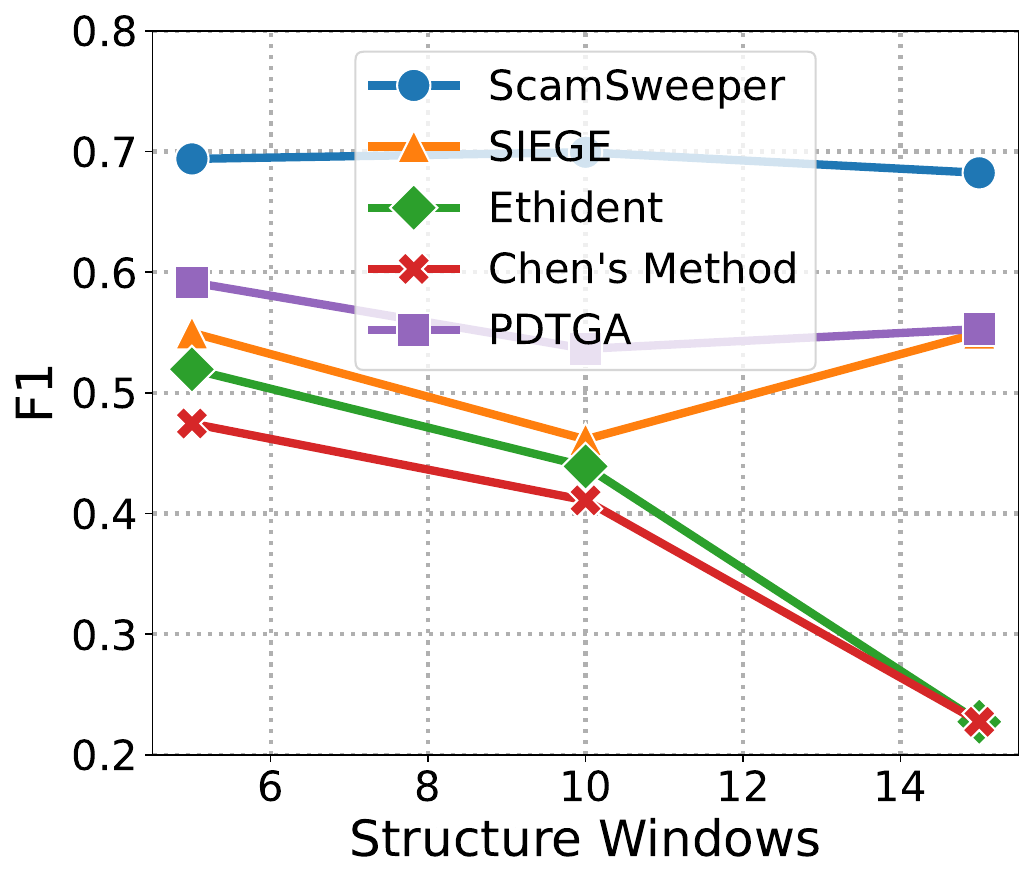}}\hspace{0.5pt}
\subfloat[\small{Precision}]{\includegraphics[width=.5\columnwidth]{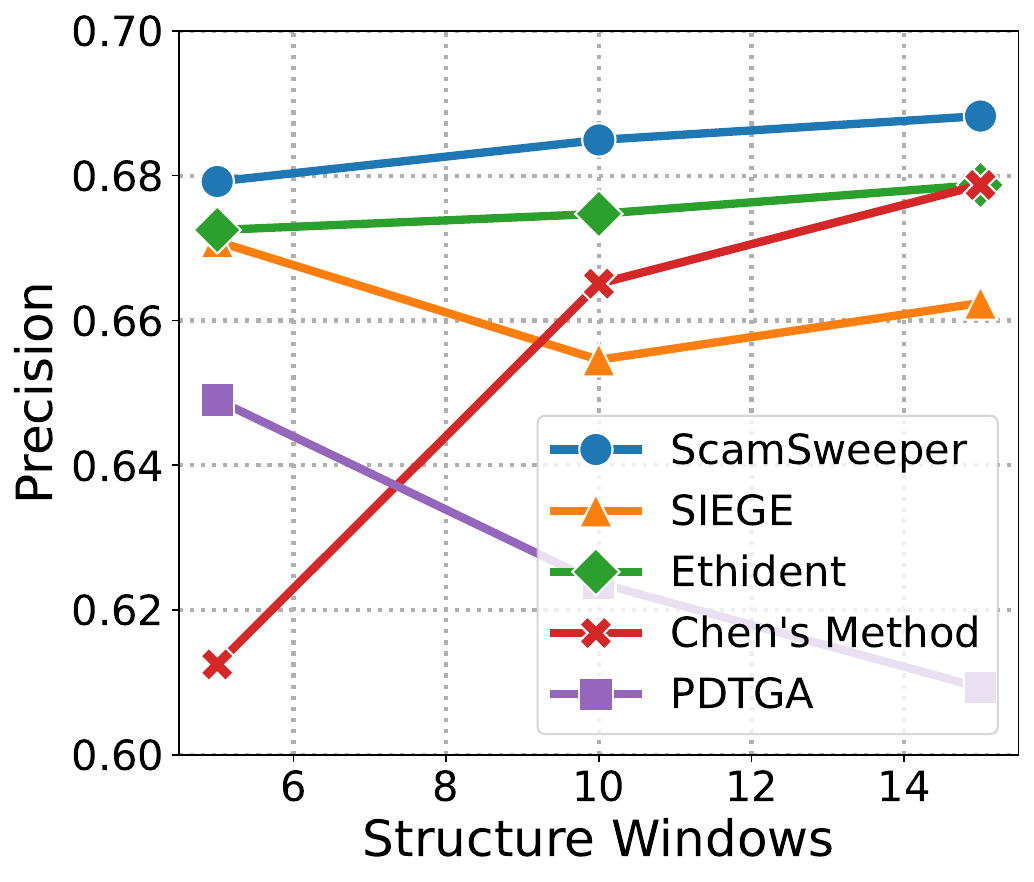}}\hspace{0.5pt}
\subfloat[\small{Recall}]{\includegraphics[width=.5\columnwidth]{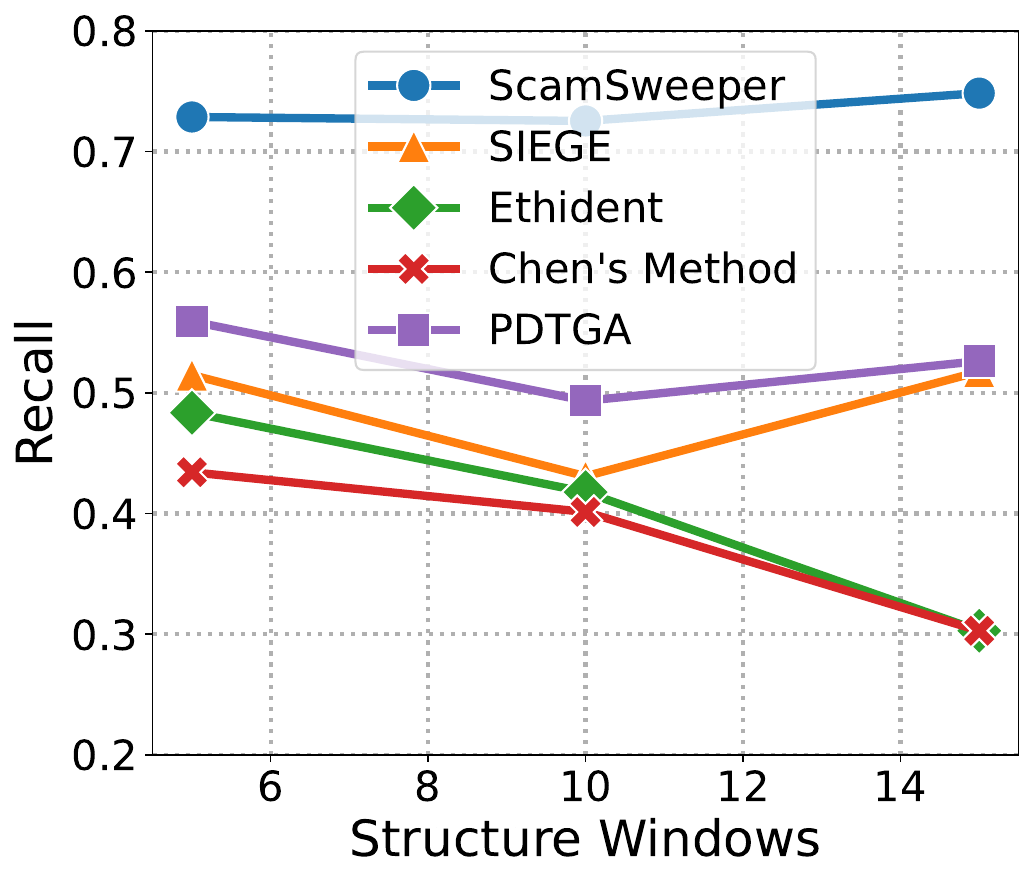}}\\
\caption{The Comparison Results on Web3 Scams Detection. The blue line represents the ScamSweeper, the orange line indicates the SIEGE, the green line means the Ethident, the red line is Chen's method, and the purple line is the PDTGA.}
\vspace{-2ex}
\label{fig:web3comparison}
\end{figure*}

In Fig. \ref{fig:Frequency_Distribution}, we can observe that the area of the curve, in terms of its horizontal and vertical coordinates, is represented in total degrees. The larger the area between the orange and blue curves in the graph, the greater the difference between the discrepancies. Therefore, the empirical data indicates that only a restricted number of pivotal nodes are accountable for initiating transactions in web3 scams and phishing networks. However, the number of receiving nodes in phishing is higher than in the web3 scam network, so the profit nodes of the web3 scam network are more concentrated.

Therefore, the data suggests that the web3 scams network has a higher propensity to expand outward (out-degree) and a lower inclination to expand inward (in-degree) than the phishing network. So, the directed feature learning of graph structure in ScamSweeper can match the in-out degree attribute features and distinguish normal nodes from malicious nodes. 
The limited number of nodes will produce a multitude of transactions, resulting in introducing a lot of noise by the methods that only consider the graph structure.
Therefore, to effectively analyze the dynamic evolution in transaction data and to differentiate between normal nodes and illegal nodes (i.e., phishing and web3 scam accounts), the sequence information of the graph structure should be considered.

\begin{tcolorbox}[boxrule=1pt,boxsep=1pt,left=3pt,right=3pt,top=3pt,bottom=3pt]
\textbf{Answer to RQ1.}
The degree-frequency follows a power-law distribution, indicating that only a few nodes are crucial in the entire network. The key feature is that the three nodes possess unique in- and out-degree distributions. 

\end{tcolorbox}

\subsection{RQ2: Detection on Web3 Scam Accounts} \label{subsec:RQ3}

To validate the effectiveness of ScamSweeper in detecting web3 scams and answer RQ2, we conduct experiments with various graph learning techniques (i.e., the method utilized by Chen et al.~\cite{chenliang2020phishing}, Ethident proposed by Zhou et al.~\cite{zhou2022behavior}, and SIEGE presented by Li et al.~\cite{li2023siege}), and the sequence learning method (i.e., PDTGA introduced by Wang et al.~\cite{wang2023phishing}).

We compare them with ScamSweeper across different structure window sizes, within the 3,125 labeled Web3 scams dataset described in Table \ref{tab:dataset_analysis}. Specifically, to evaluate it, we randomly select the same number of nodes from the network other than web3 scams as negative samples, balancing the dataset. Furthermore, we implement a three-layer structure for GAT, GCN, and GraphSAGE in Ethident, Chen's method, and SIEGE. GraphSAGE is a type of GNN model that samples a fixed number of neighbor nodes, and we employ a three-layer structure with mean aggregation. The PDTGA method has been enriched with several key components, including the self-attention, feedforward neural network, mask mechanism, and position embedding. Within it, the hidden dimension size has been set to 16, which is the same as ScamSweeper.

As illustrated in Fig. \ref{fig:web3comparison}, our experimental results show that ScamSweeper performs better than other methods. As for graph methods, at a structure window of 15, ScamSweeper scores a recall of 0.75, which is 0.23-0.44 higher than other graph methods. Therefore, ScamSweeper achieves higher F1-scores than other methods, with scores of 0.69, 0.70, and 0.68 in the three structure windows. In comparison, the highest F1-scores achieved by Ethident, Chen's method, and SIEGE are 0.52, 0.47, and 0.55, respectively. Note that the three indicators used in RQ2 are weighted for positive and negative samples, thus providing an accurate reflection to distinguish between the malicious and normal nodes. The PDTGA achieves better F1-score and recall than graph learning methods, but there is still a certain gap compared with ScamSweeper. As shown in Fig. \ref{fig:web3comparison}, for [F1-score, precision, recall], ScamSweeper outperforms PDTGA by at least about [17.29\%, 4.64\%, 30.29\%].

\begin{tcolorbox}[boxrule=1pt,boxsep=1pt,left=3pt,right=3pt,top=3pt,bottom=3pt]
\textbf{Answer to RQ2.}
When the structural window sets 10, ScamSweeper achieves a weighted F1-score of 0.70, significantly 17.29\%-48.94\% higher than other methods. 

\end{tcolorbox} 

\subsection{RQ3: Impact of Each Component} \label{subsec:RQ2}
To show the efficiency of the components (\cref{sub:graph_construction}, \cref{subsec:graph_encoder}, and \cref{subsec:sequence_learning}), we conduct an ablation experiment.

We carry out an ablation study, with the final classification outcomes as the results of the influence of various components. The experiment data are processed through the STRWalk, acquiring graphs from a large-scale network (See \cref{sub:graph_construction}). Subsequently, experiments are conducted on the graphs, enabling further comparison. In the subgraph learning stage (See \cref{subsec:graph_encoder}), we directly remove the graph-learning layer for comparison, illustrating the impact of graph-learning techniques within ScamSweeper. In the subgraph sequence learning phase (See \cref{subsec:sequence_learning}), we substitute the transposed Transformer encoder with a conventional Transformer structure, e.g., the different structure depicted in Fig. \ref{fig:t-Transformer}, indicating the influence of the transposed Transformer. Specifically, we set the number of encoder layers to 1, the number of heads to 2, the feature size to 16, the temporal interval to 7 days, and the temporal length to 3 years. Notably, we focus on evaluating the ability of each component in web3 scam detection.

\begin{table}[ht]
\centering
  \caption{The Ablation Results of Each Component. The @ indicates the structure window in the STRWalk. The ScamSweeper-t is without the transposed Transformer, and the ScamSweeper-g is without the graph learning.}
  \vspace{-1ex}
  \label{tab:F1-ablation}
  \begin{tabular}{c c c c c} 
    \toprule
    \textbf{Metric} & \textbf{Method} & \textbf{@5} & \textbf{@10} & \textbf{@15} \\
    \midrule 
     \multirow{3}{*}{F1-score}  & ScamSweeper-t &  0.41  &  0.43  &  0.44  \\
     &  ScamSweeper-g &  0.53 & 0.51  &  0.55  \\
     &  \textbf{ScamSweeper}  &  \textbf{0.78} & \textbf{0.78}  & \textbf{0.73}   \\
\midrule 
     \multirow{3}{*}{\makecell[c]{Weighted\\F1-score}} & ScamSweeper-t &  0.42  &  0.46  &  0.47  \\
    &      ScamSweeper-g & 0.57  &  0.55 &  0.58  \\
    &   \textbf{ScamSweeper}   & \textbf{0.69}  &  \textbf{0.70}  &  \textbf{0.68}  \\
  \bottomrule 
\end{tabular}
\end{table}

Upon thorough examination of all three structural windows shown in Table \ref{tab:F1-ablation}, the ScamSweeper-t in Table \ref{tab:F1-ablation} shows a range of positive F1-scores of 0.41-0.44 at window size 5. This range is comparatively lower than ScamSweeper-g, which records a range of 0.51-0.55. However, when the window size is 15, ScamSweeper's ability to detect positive samples is weaker. The reason is that the window, i.e., temporal interval, contains more noisy data, resulting in a relatively chaotic feature matrix after the phase of the graph learning encoder. However, the enhanced performance of ScamSweeper-g is observed in detecting scams when the structure window is 15 from 10, which suggests that the transposed Transformer can alleviate the issue by focusing on the dynamic evolution of all the temporal features in the embedding dimension. Thus, the impact of the graph learning on ScamSweeper is more significant than the impact of the transposed Transformer.

As shown in Table \ref{tab:F1-ablation}, ScamSweeper is top-performing across all three structural windows compared to other incomplete scenarios. ScamSweeper achieved weighted F1-scores of \{0.69, 0.70, 0.68\} at the three structural windows, which is 0.1-0.15 higher than ScamSweeper-g and 0.21 to 0.27 higher than ScamSweeper-t. However, when the window size is 10, there is a decrease in the web3 scams detection capacity of ScamSweeper. Nonetheless, ScamSweeper's weighted F1-score is higher than others, indicating superior performance in identifying web3 scams and normal nodes. 

Drawing from the experiment's findings, it may be inferred that the comparison results retain the order of ScamSweeper $>$ only sequence learning $>$
only graph learning, even when the STRWalk is restarted. Additionally, Table \ref{tab:F1-ablation} reveals that there is a minimal variation (up to 0.05 in the F1-score and 0.02 in the weighted F1-score) among the three times of STRWalk. It suggests that the randomness of STRWalk has minimal influence on the performance of ScamSweeper.

\begin{tcolorbox}[boxrule=1pt,boxsep=2pt,left=3pt,right=3pt,top=3pt,bottom=3pt]
\textbf{Answer to RQ3.}
Graph and sequence learning components improve the F1-scores by 47.17\% and 90.24\%, respectively, with a structural window size of 5. 
Moreover, ScamSweeper has good robustness in different structural window sizes.
\end{tcolorbox}

\subsection{RQ4: Detection on Phishing Accounts} \label{subsec:RQ4}

ScamSweeper can also identify phishing scams, showcasing its generalization capabilities. Therefore, phishing nodes are assessed to evaluate the model's performance.

\begin{table}[ht]
\centering
  \caption{The Comparison Results on Phishing Node Feature Detection. \dag\space, \ddag\space, $_{dnn}$, and $_{rf}$ refer to the utilization of STRWalk, TRWalk, DNN, and random forest, respectively.}
  \vspace{-1ex}
  \label{tab:ablation_analysis1}
  \begin{tabular}{c c c c c} 
    \toprule
    \textbf{Method} & \textbf{Precision} & \textbf{Recall} & \textbf{F1-score} \\
    \midrule
    Yuan's method$^\dag$~\cite{yuan2020detecting}            &  0.82  &  0.77  &  0.79  \\
    DGTSD$^\dag$~\cite{choi2024learning}           &  0.82  &  0.79  &  0.80  \\
    BERT4ETH$^\dag$$_{dnn}$~\cite{hu2023bert4eth}           &  0.76  &  0.74  &  0.75  \\
    BERT4ETH$^\dag$$_{rf}$~\cite{hu2023bert4eth}            &  0.81  &  0.64  &  0.71  \\
    \textbf{ScamSweeper$^\ddag$} & \textbf{0.86}  &  \textbf{0.84}  &  \textbf{0.85}  \\
    \textbf{ScamSweeper$^\dag$} & \textbf{0.95}  &  \textbf{0.92}  &  \textbf{0.94}  \\
  \bottomrule 
\end{tabular}
\end{table}

Since the STRWalk is divided into two steps: time and structure sampling, and only the end of time sampling is followed by structure sampling. So, we sample the network using Node2vec, DeepWalk, temporal random walk (TRWalk), and STRWalk, respectively. Moreover, to compare the effects of these walking methods at their optimal states, we follow the parameter settings at \cite{lin2020modeling}, which compared various walking methods. Therefore, the walk length is set to 20, the window size is 4, and the embedding dimension of the nodes is 128. The max sequence length in BERT4ETH is set to 100. In addition, our data collected in Table \ref{tab:dataset_analysis} shows a significant imbalance between phishing and normal nodes, with a ratio of 4905:636. Therefore, to ensure a practical evaluation of phishing node detection without introducing noise, we utilize the dataset called \texttt{EPTransNet}~\cite{EPTransNet2024}, which is a subset of our dataset containing 1,165 phishing nodes. We then supplement it with the 636 normal nodes we collected to address the imbalance of positive and negative samples in our dataset. 

As shown in Table \ref{tab:ablation_analysis1}, the TRWalk with only time and the STRWalk with time and structure in ScamSweeper can achieve better performance than existing methods. We can find that ScamSweeper with STRWalk can achieve over 17.5\%, 15.85\%, and 16.46\% higher than other methods in F1-score, precision, and recall. ScamSweeper with TRWalk can achieve 6.25\%, 4.88\%, and 6.33\% higher than the other methods in terms of F1-score, precision, and recall. The original Node2Vec and DeepWalk in Yuan's method and DGTSD both use the random walk to sample the network. We only show the case when Yuan's method and DGTSD use STRWalk. For further comparison, we describe the benefits of random walk and STRWalk for Yuan's method and DGTSD in \cref{subsec:strwalk_optimation}.

\begin{figure}[!t]
\vspace{-2ex}
\centering
\subfloat[\footnotesize{Yuan's method$^\dag$}]{\includegraphics[width=.31\columnwidth]{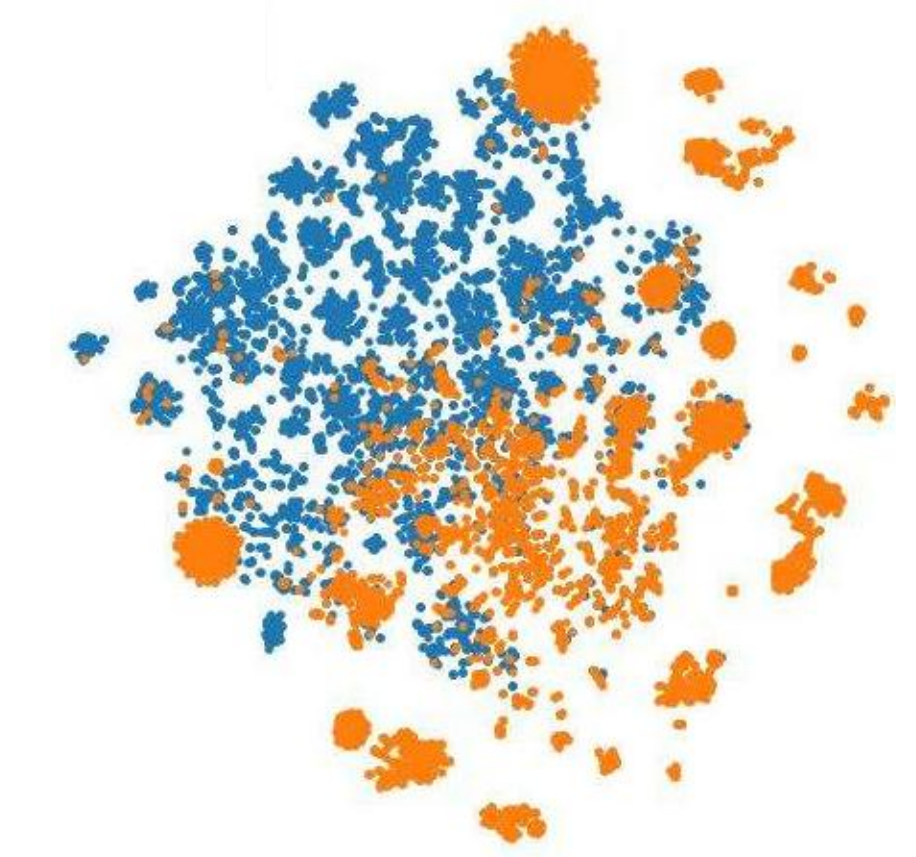}}\hspace{0.5pt}
\subfloat[\footnotesize{DGTSD$^\dag$}]{\includegraphics[width=.31\columnwidth]{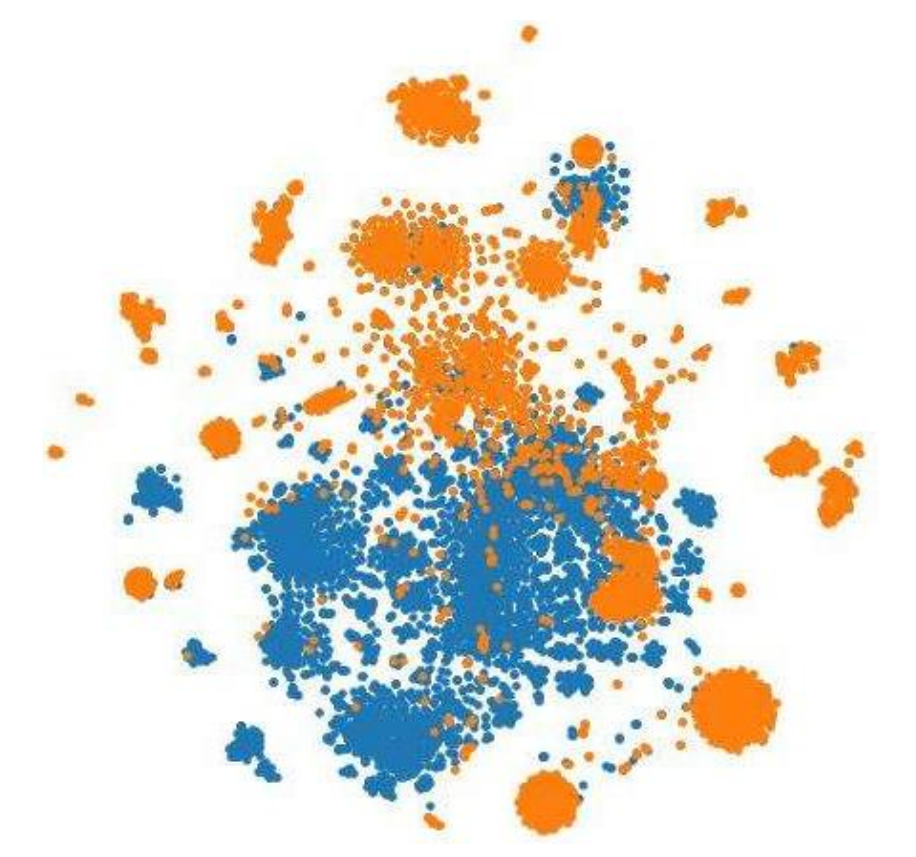}}\hspace{0.5pt}
\subfloat[\footnotesize{ScamSweeper$^\ddag$}]{\includegraphics[width=.31\columnwidth]{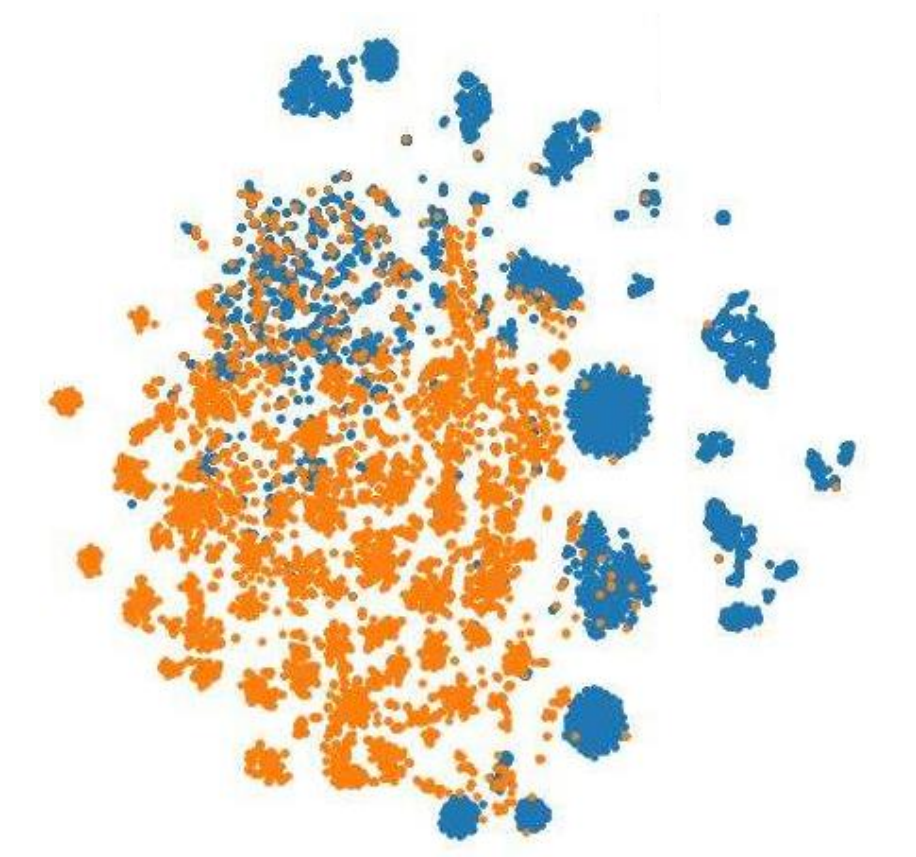}}\\ \vspace{-2ex}
\subfloat[\footnotesize{BERT4ETH$^\dag$$_{dnn}$}]{\includegraphics[width=.31\columnwidth]{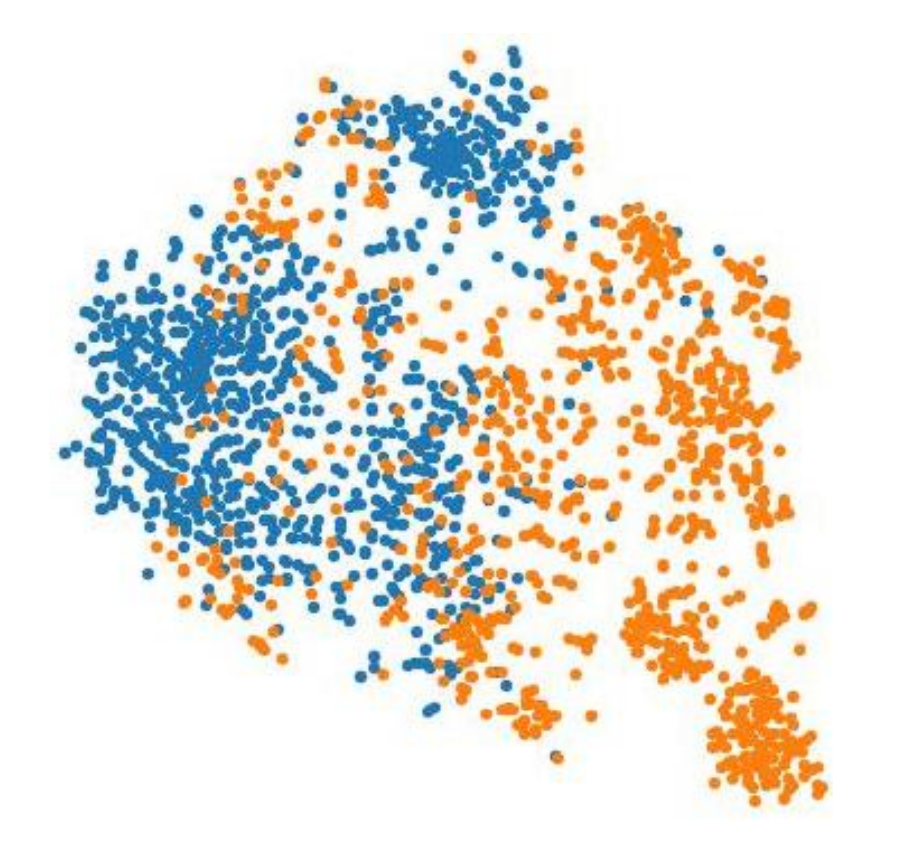}}\hspace{0.5pt}
\subfloat[\footnotesize{BERT4ETH$^\dag$$_{rf}$}]{\includegraphics[width=.31\columnwidth]{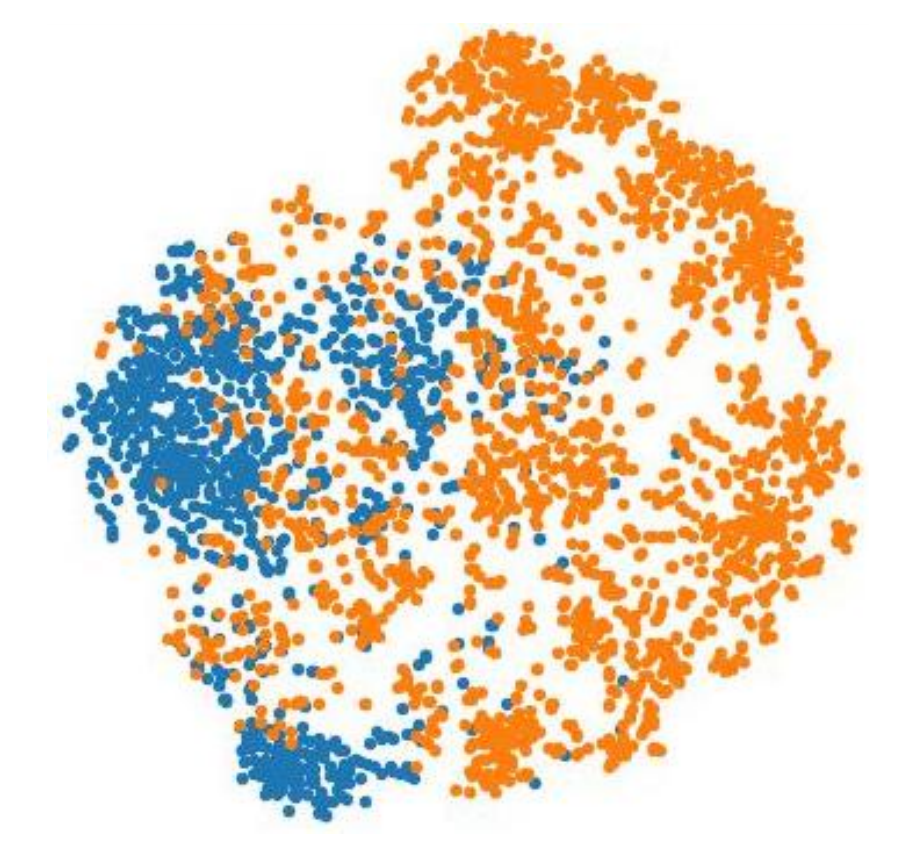}}\hspace{0.5pt}
\subfloat[\footnotesize{ScamSweeper$^\dag$}]{\includegraphics[width=.31\columnwidth]{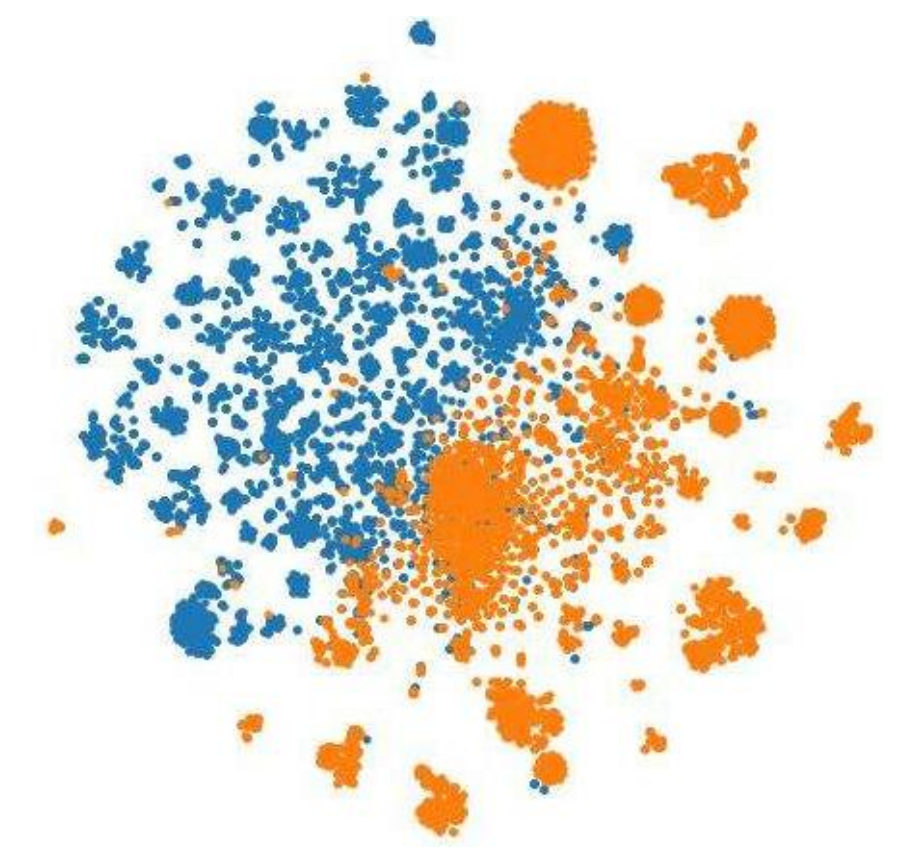}}\\
\vspace{-1ex}
\caption{The Embedding Visualization of Node Feature Distribution by T-SNE, where \colorbox{orange}{\phantom{}} represents the malicious accounts, and \colorbox{blue}{\phantom{}} means the normal accounts.}
\label{fig:walks_Comparison}
\vspace{-4ex}
\end{figure}

To visualize the distribution of classified nodes, we leverage t-SNE~\cite{van2008visualizing} as the classification algorithm, representing normal and malicious nodes in blue and orange, respectively. Since the number of negative nodes in the dataset is larger than the positive nodes, we sample the same number of node features from the negatives.
Fig. \ref{fig:walks_Comparison} shows the performance of different models on phishing detection, i.e., Yuan's method, DGTSD, TRWalk, BERT4ETH$_{dnn}$, BERT4ETH$_{rf}$, and STRWalk. The BERT4ETH$_{dnn}$ and  BERT4ETH$_{rf}$ utilize the DNN~\cite{chen2023dycl,hu2023bert4eth} or random forest~\cite{hu2023bert4eth} algorithm to identify phishing nodes.
ScamSweeper$\dag$ refers to the phenomenon that utilizes the STRWalk. It can be found that the TRWalk demonstrates superior discriminative capability compared to Yuan's method and DGTSD, thus confirming the enhanced effectiveness of temporal attributes in detecting malicious nodes within networks. Furthermore, the ScamSweeper$\dag$ with STRWalk are separable, and most malicious nodes can be linearly separated.

\begin{tcolorbox}[boxrule=1pt,boxsep=2pt,left=3pt,right=3pt,top=3pt,bottom=3pt]
\textbf{Answer to RQ4.}
ScamSweeper with STRWalk can identify phishing nodes to a certain degree, resulting in an enhancement of 17.5\% in F1-score, 15.85\% in precision, and 16.46\% in recall.
\end{tcolorbox}

\section{Discussion}

\subsection{Case Study}
\label{subsec:case}
We identify a web3 scam account~\cite{case12024} based on the Drainer toolkit and a traditional phishing account~\cite{case22024} as cases, which have been reported and annotated in Etherscan~\cite{etherscan2024}.

\subsubsection{\textbf{Case I}}
\begin{figure}[ht]
\centering
\vspace{-1ex}
\subfloat[\small{Original}]{\includegraphics[width=.48\columnwidth]{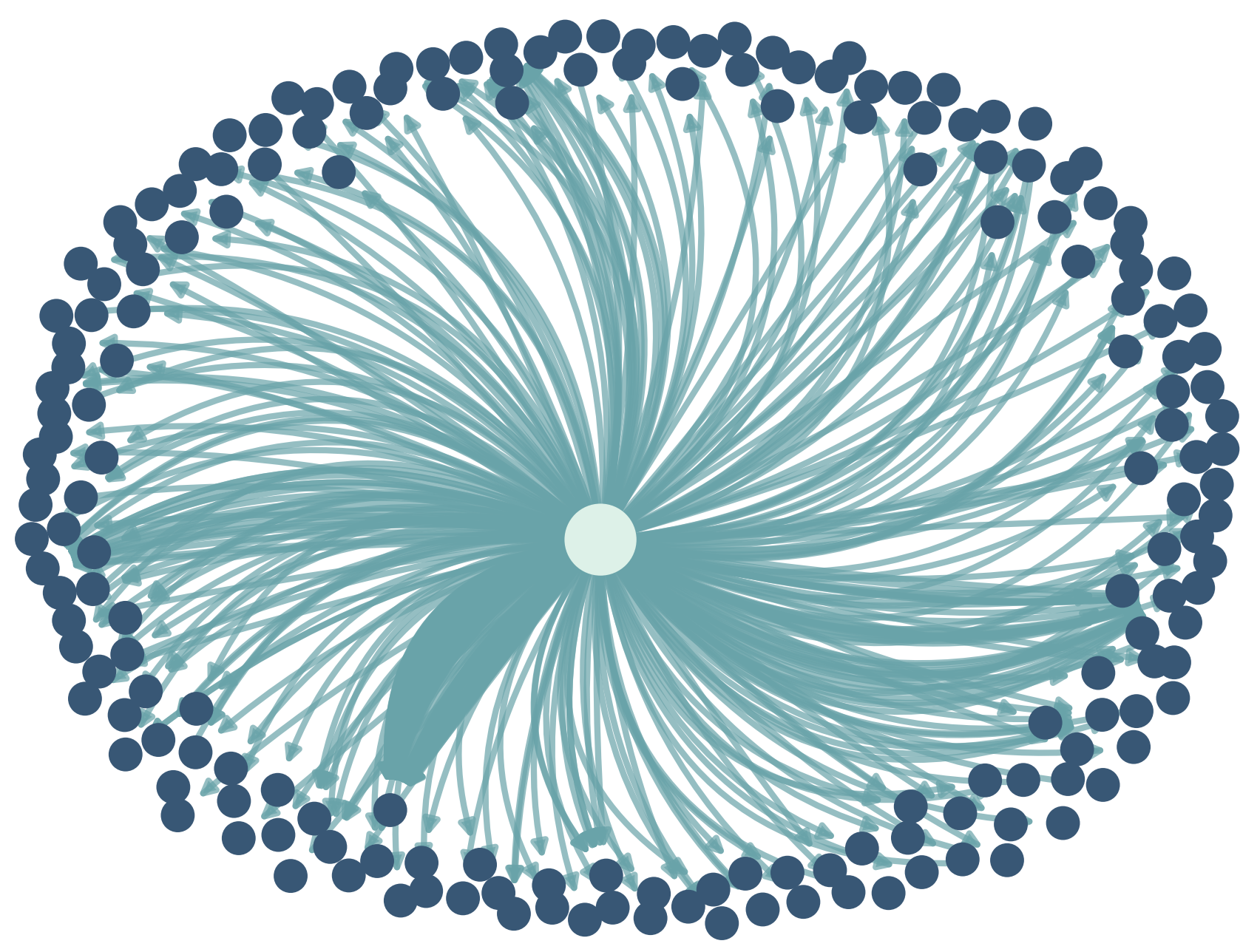}}\hspace{1pt}
\subfloat[\small{Detected Pattern}]{\includegraphics[width=.48\columnwidth]{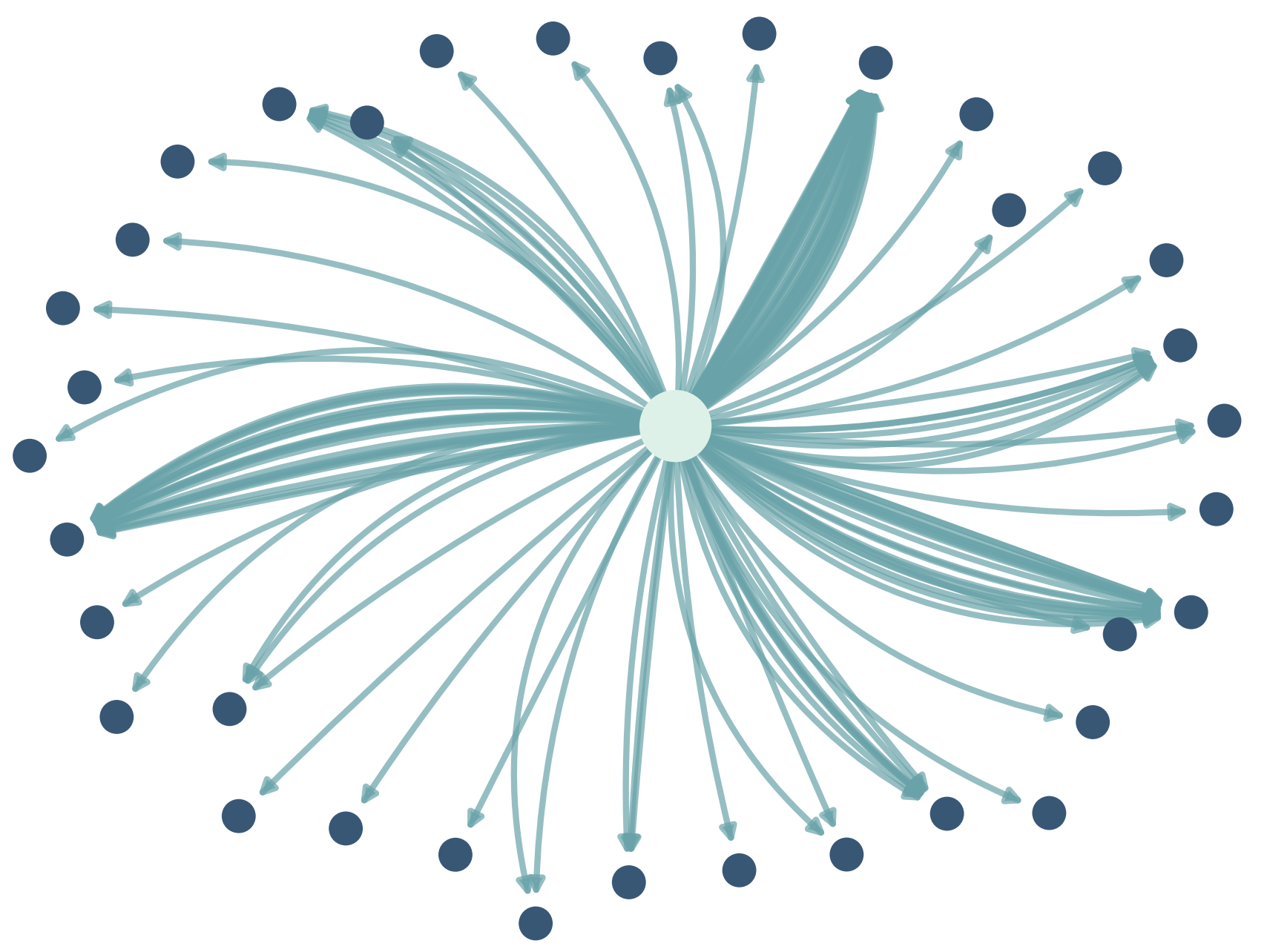}}\hspace{1pt}
\vspace{-1ex}
\caption{An Example of Transaction Network of Web3 Scam Accounts, where \colorbox{mySourceNodecolor}{\phantom{}} represents the source account of the network, and \colorbox{myOtherNodecolor}{\phantom{}} means other connected accounts. }
\vspace{-2ex}
\label{fig:Web3scams_case}
\end{figure}

\begin{figure}[ht]
\centering
\subfloat[\small{$(0,\tau)$}]{\includegraphics[width=.23\columnwidth]{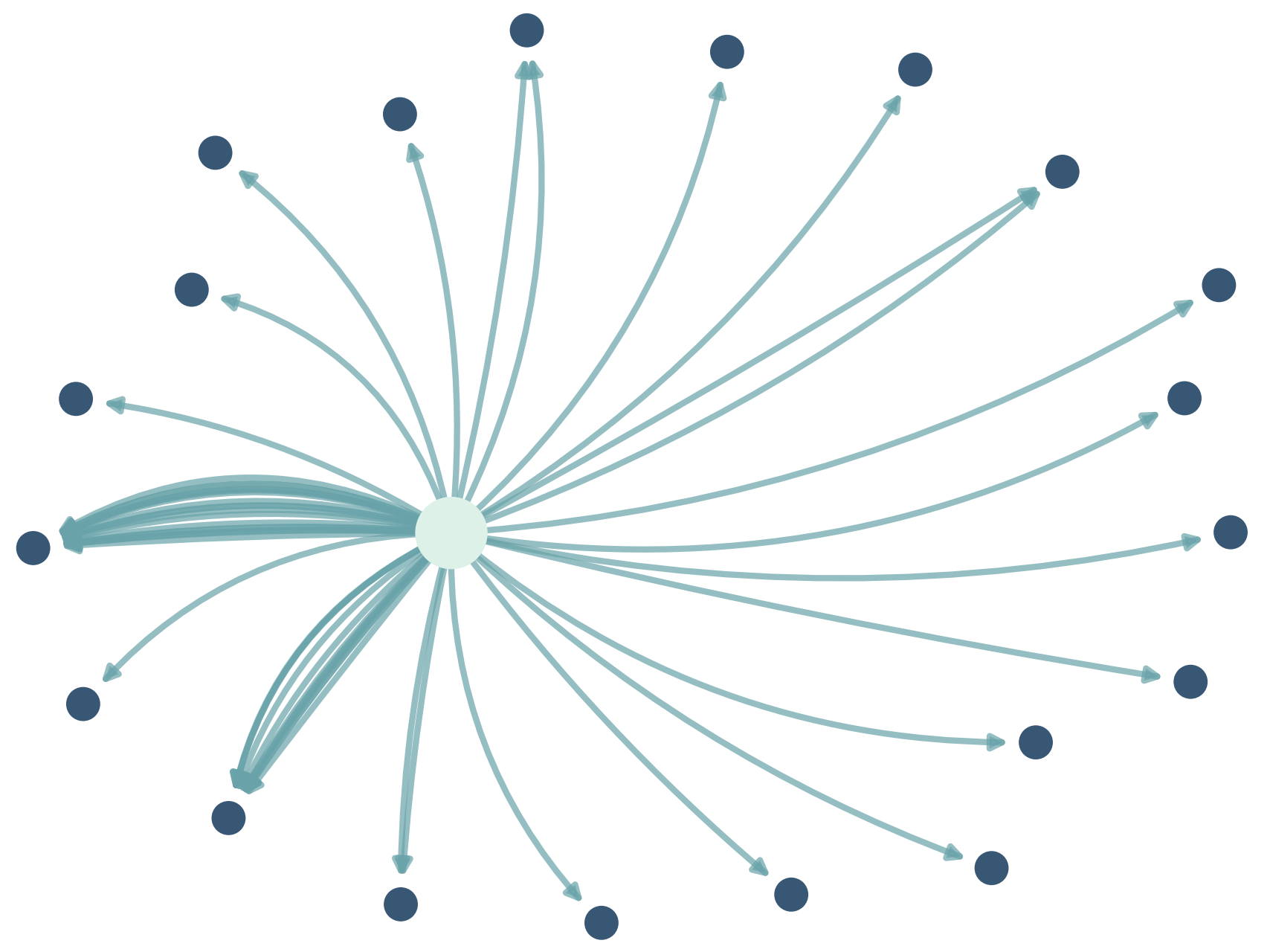}}\hspace{1pt}
\subfloat[\small{$(\tau$, $2\tau)$}]{\includegraphics[width=.23\columnwidth]{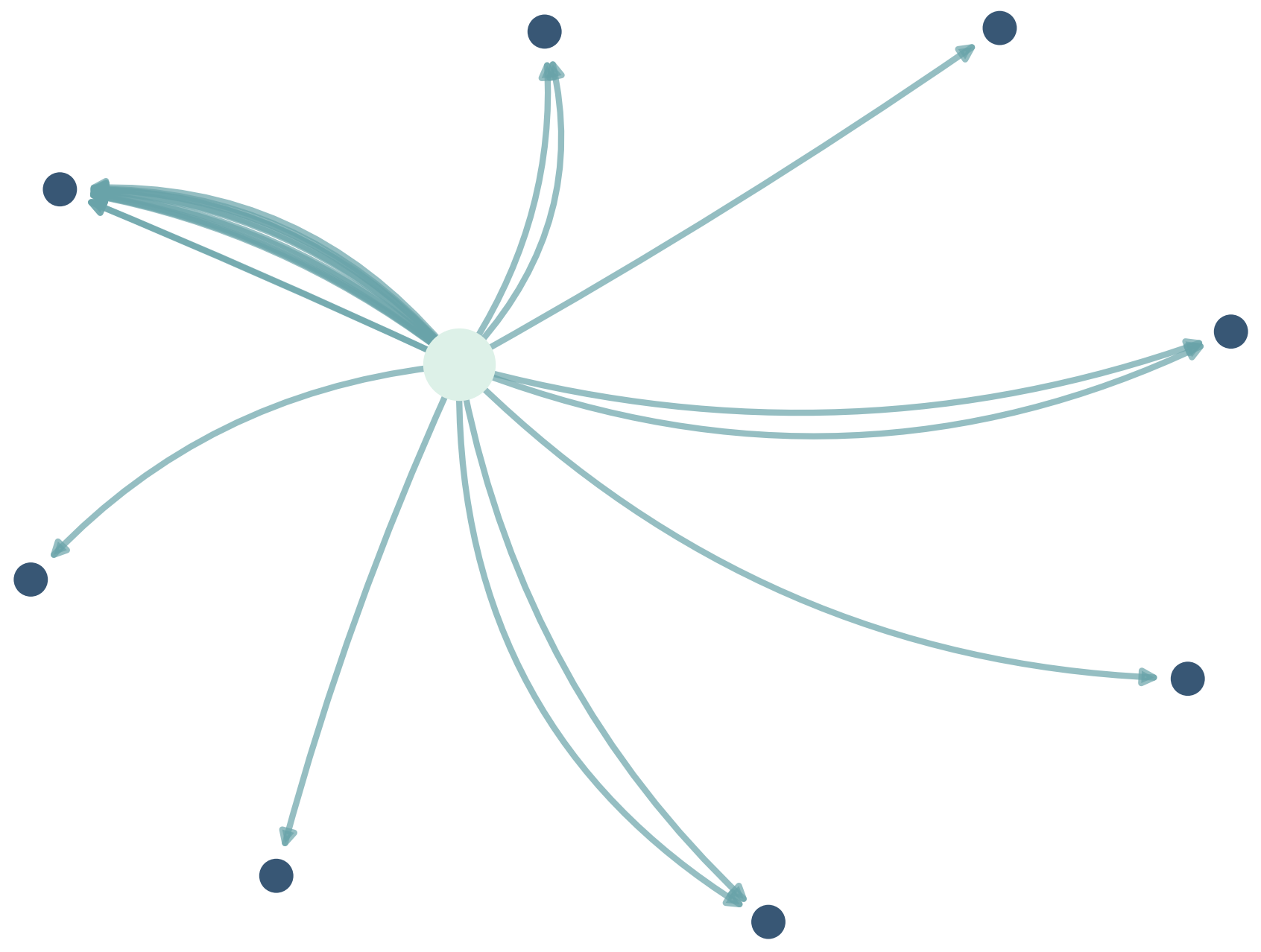}}\hspace{1pt}
\subfloat[\small{$(2\tau$,$3\tau)$}]{\includegraphics[width=.23\columnwidth]{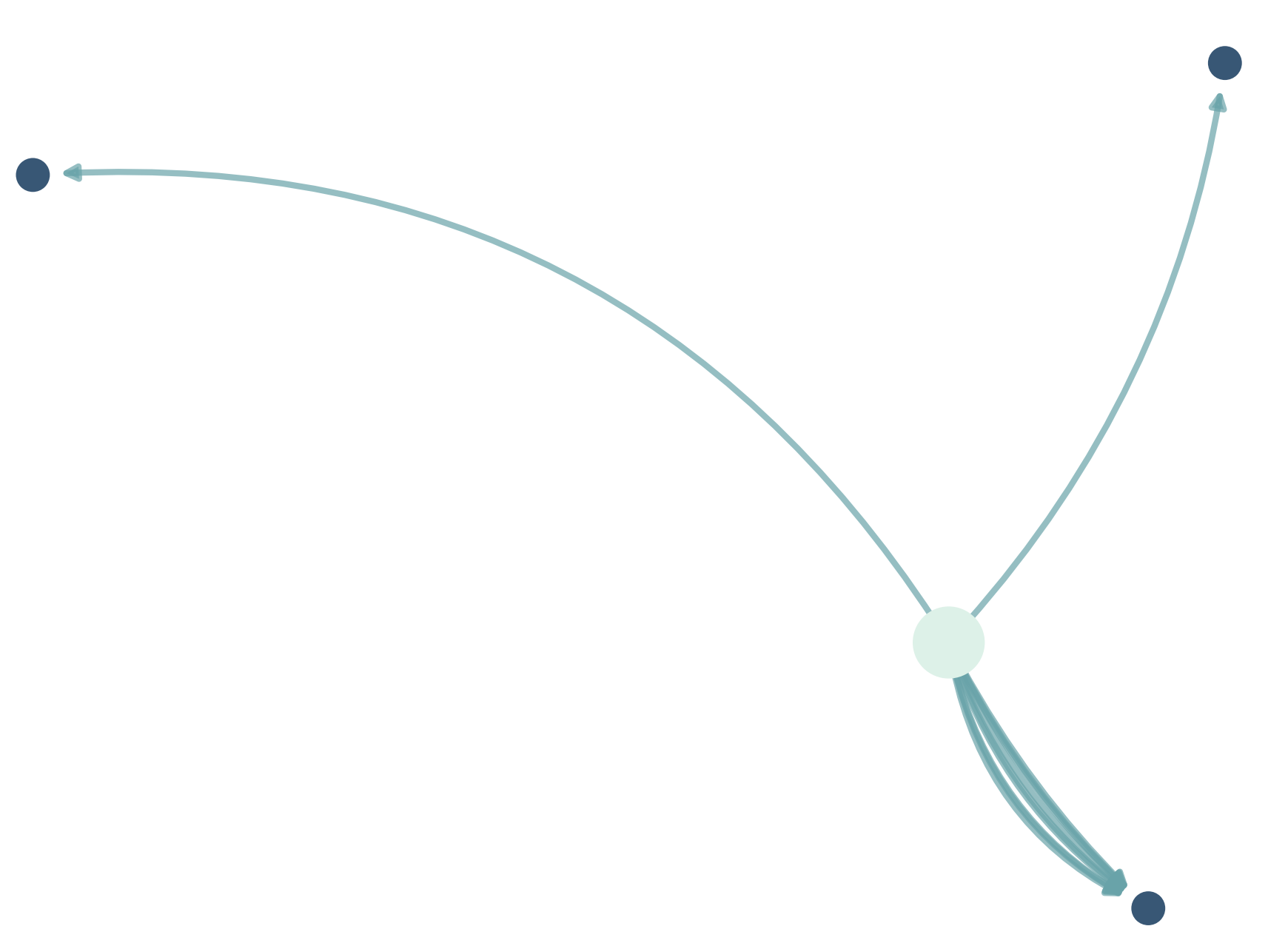}}\hspace{1pt}
\subfloat[\small{$(5\tau$, $6\tau)$}]{\includegraphics[width=.23\columnwidth]{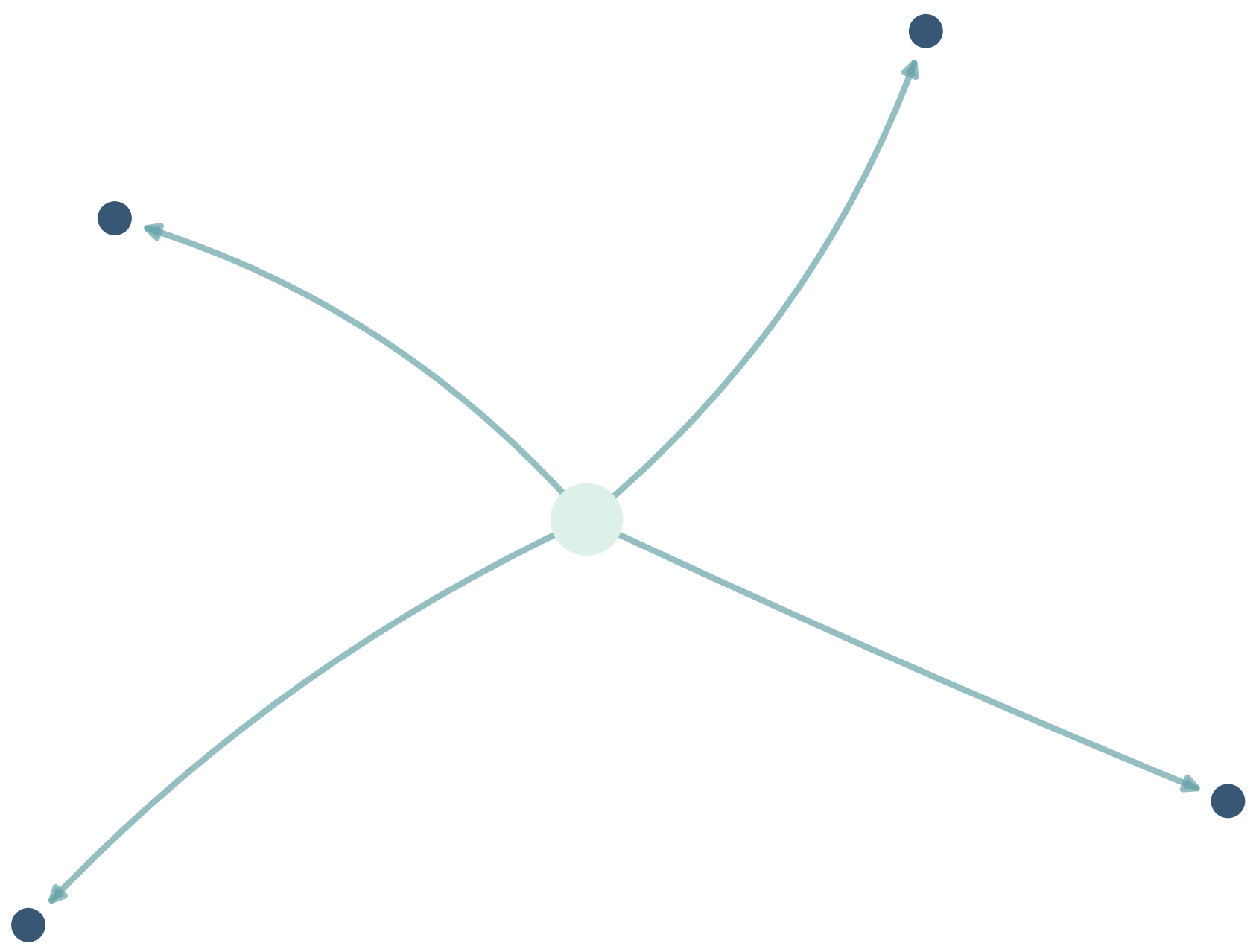}}\\ \vspace{-2ex}
\subfloat[\small{$(6\tau$, $7\tau)$}]{\includegraphics[width=.24\columnwidth]{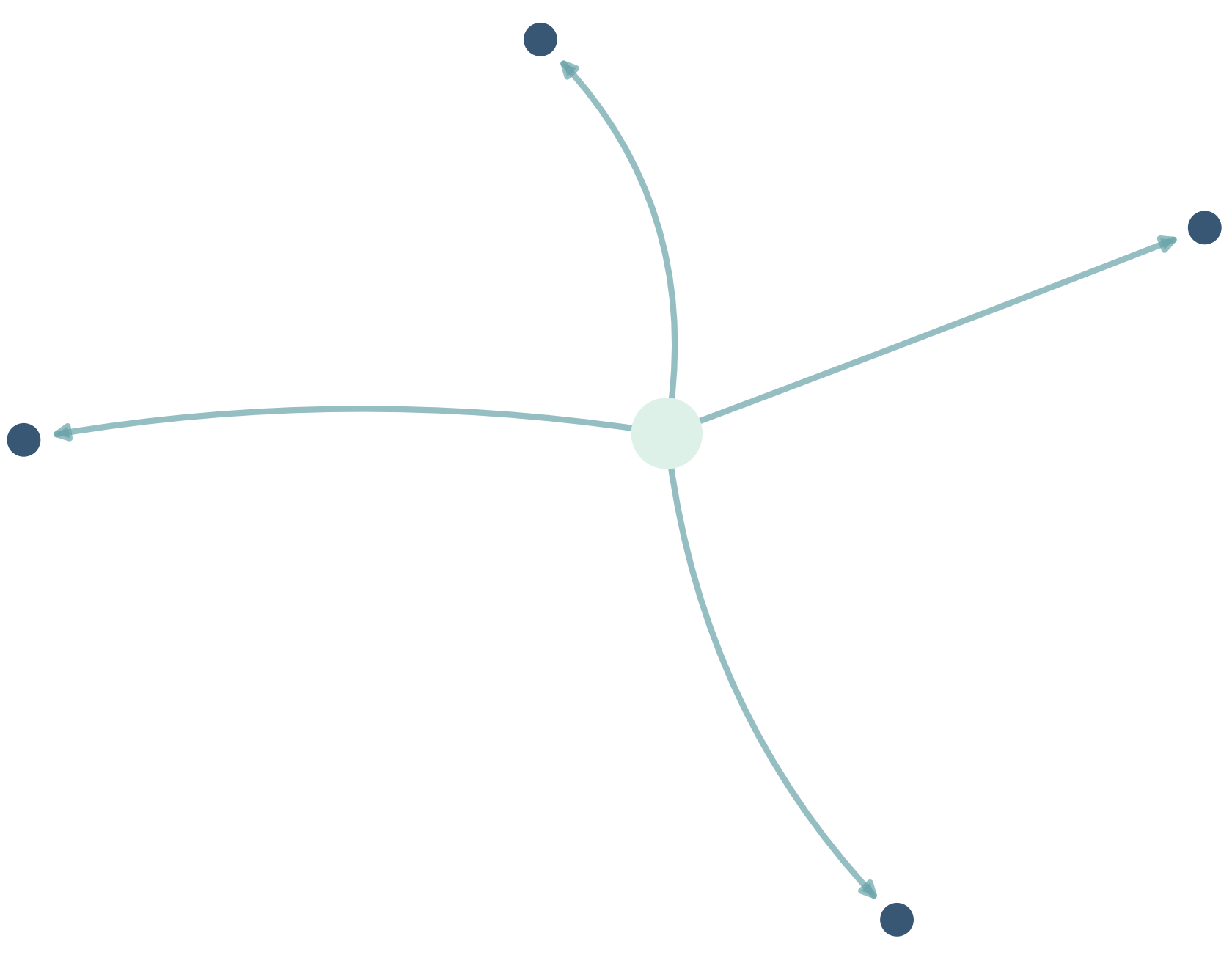}}\hspace{2pt}
\subfloat[\small{$(9\tau$, $10\tau)$}]{\includegraphics[width=.24\columnwidth]{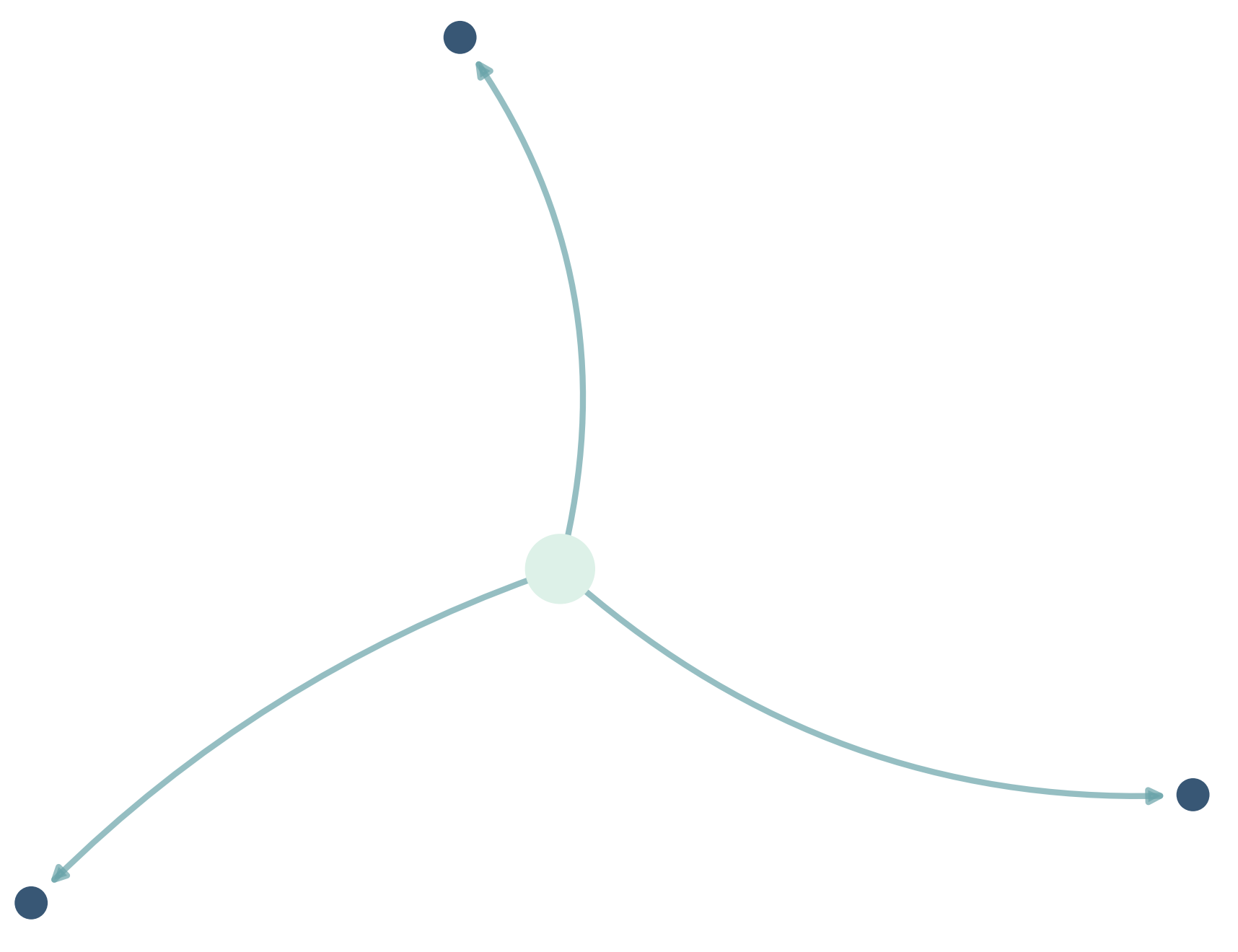}}\hspace{2pt}
\subfloat[\small{$(10\tau$, $11\tau)$}]{\includegraphics[width=.24\columnwidth]{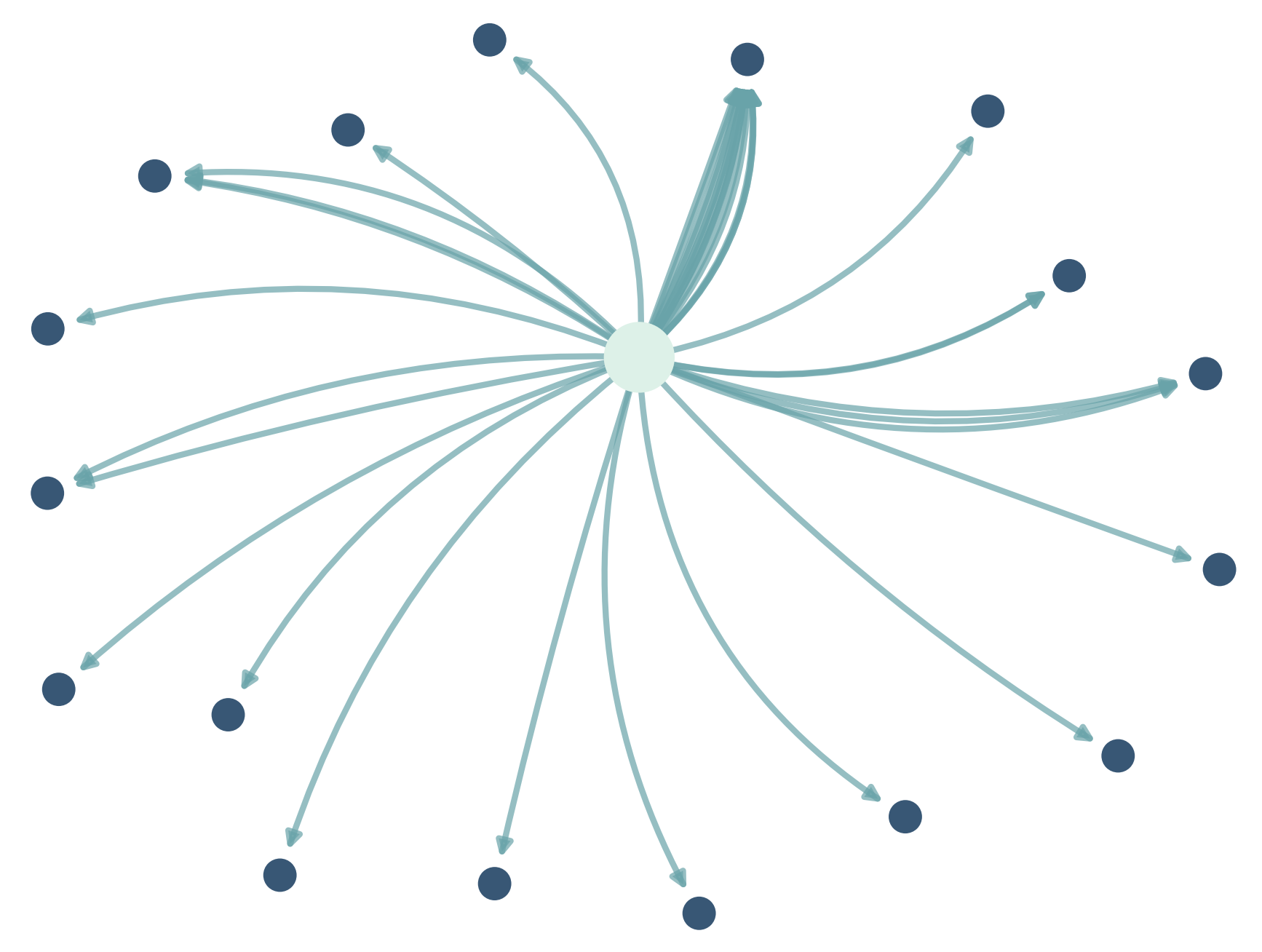}}\hspace{2pt}
\vspace{-1ex}
\caption{A Dynamic Evolution Example of Transaction Network of the Mimic Account in the Web3 Scam as Service. }
\label{fig:Web3scams_case_dynamicEvolution}
\end{figure}

The accounts serving web3 services send a large number of transaction requests. These attackers utilize platforms such as service providers to promote fraudulent information in large quantities. Fig. \ref{fig:Web3scams_case}(a) shows an example of the transaction graph of a scam attacker that provides web3 services, where the length of the edges is inversely proportional to their timestamp. The Fig. \ref{fig:Web3scams_case}(b) is the transaction graph sampled by the STRWalk. We set the value of $\tau$ to 7 days, and the structure window is set to be unrestricted. The dynamic evolutions are shown in Fig. \ref{fig:Web3scams_case_dynamicEvolution}. We can observe that the number of neighbor nodes gradually decreases with time. Finally, it diffuses to a large number of accounts in the last time interval.

The fraudulent activity commenced by mimicking typical transaction patterns initially during ($0,\tau$) and then gradually progressed to defrauding users during ($\tau,10\tau$). Throughout this phase, the transactions have been focused on fewer accounts to transfer profits into some individual accounts. After profiting, a fresh round of mimicking during ($10\tau,11\tau$) could have commenced.

\subsubsection{\textbf{Case II}}
\begin{figure}[ht]
\centering
\subfloat[\small{$(0,\tau)$}]{\includegraphics[width=.23\columnwidth]{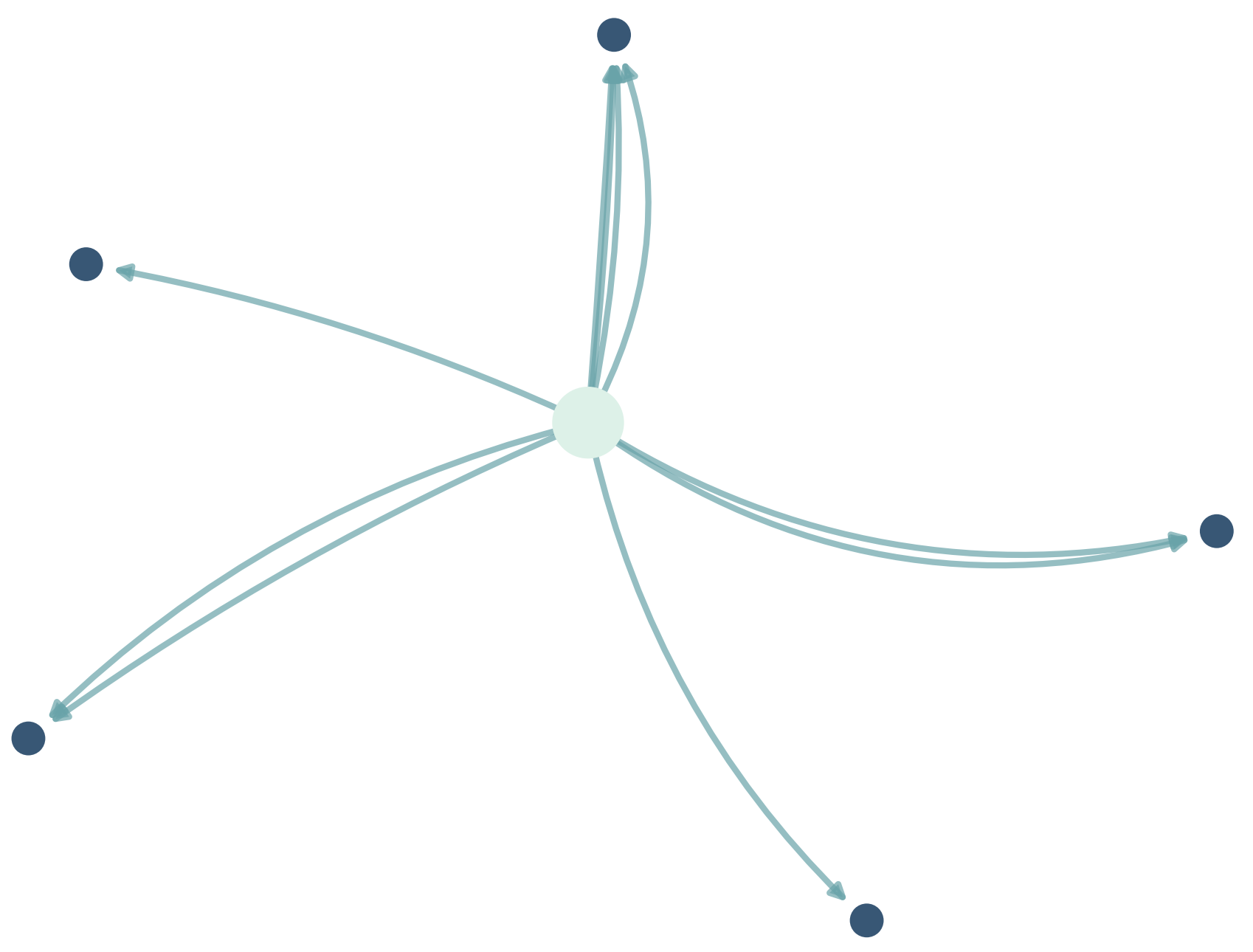}}\hspace{1pt}
\subfloat[\small{$(\tau$, $2\tau)$}]{\includegraphics[width=.23\columnwidth]{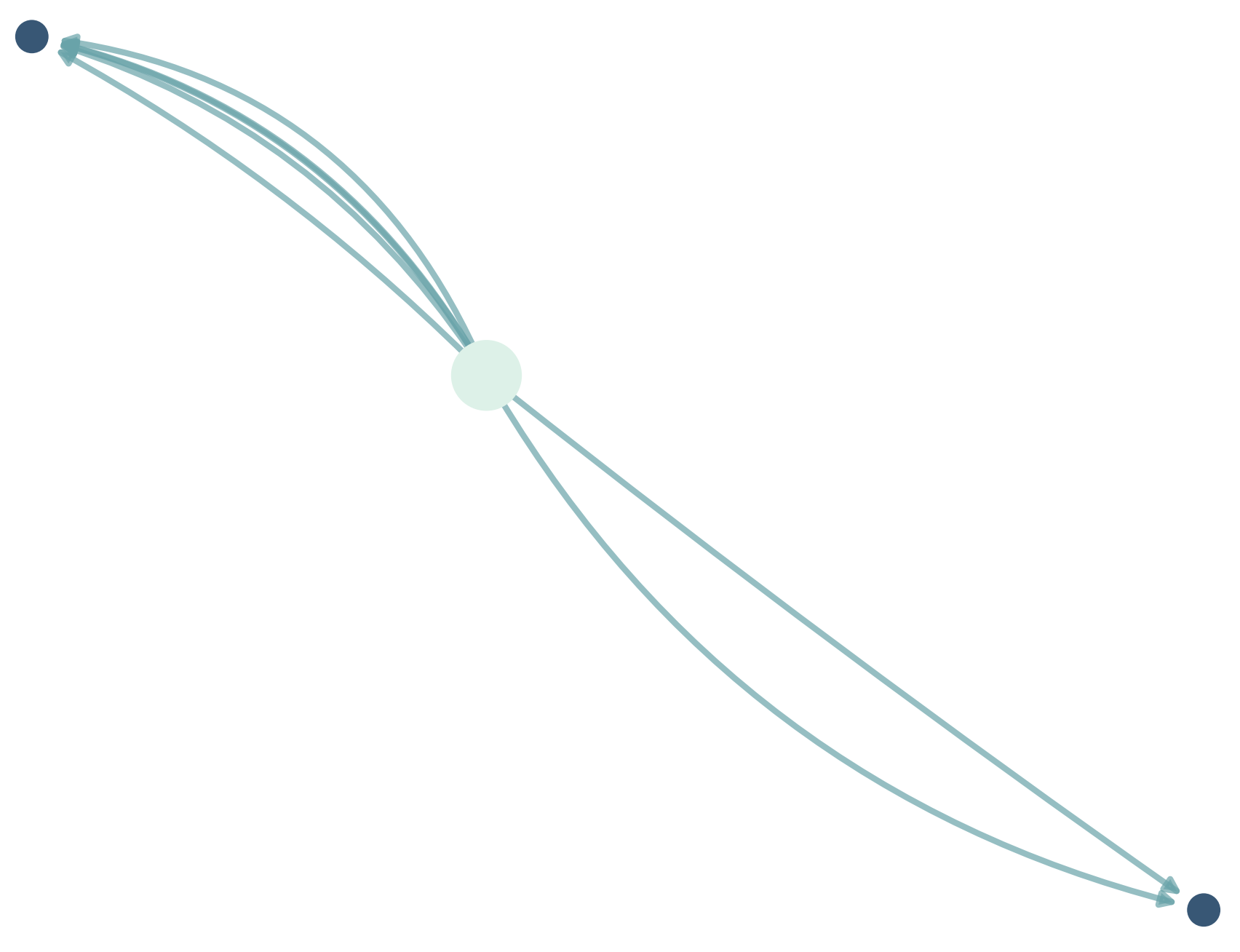}}\hspace{1pt}
\subfloat[\small{$(3\tau$, $4\tau)$}]{\includegraphics[width=.23\columnwidth]{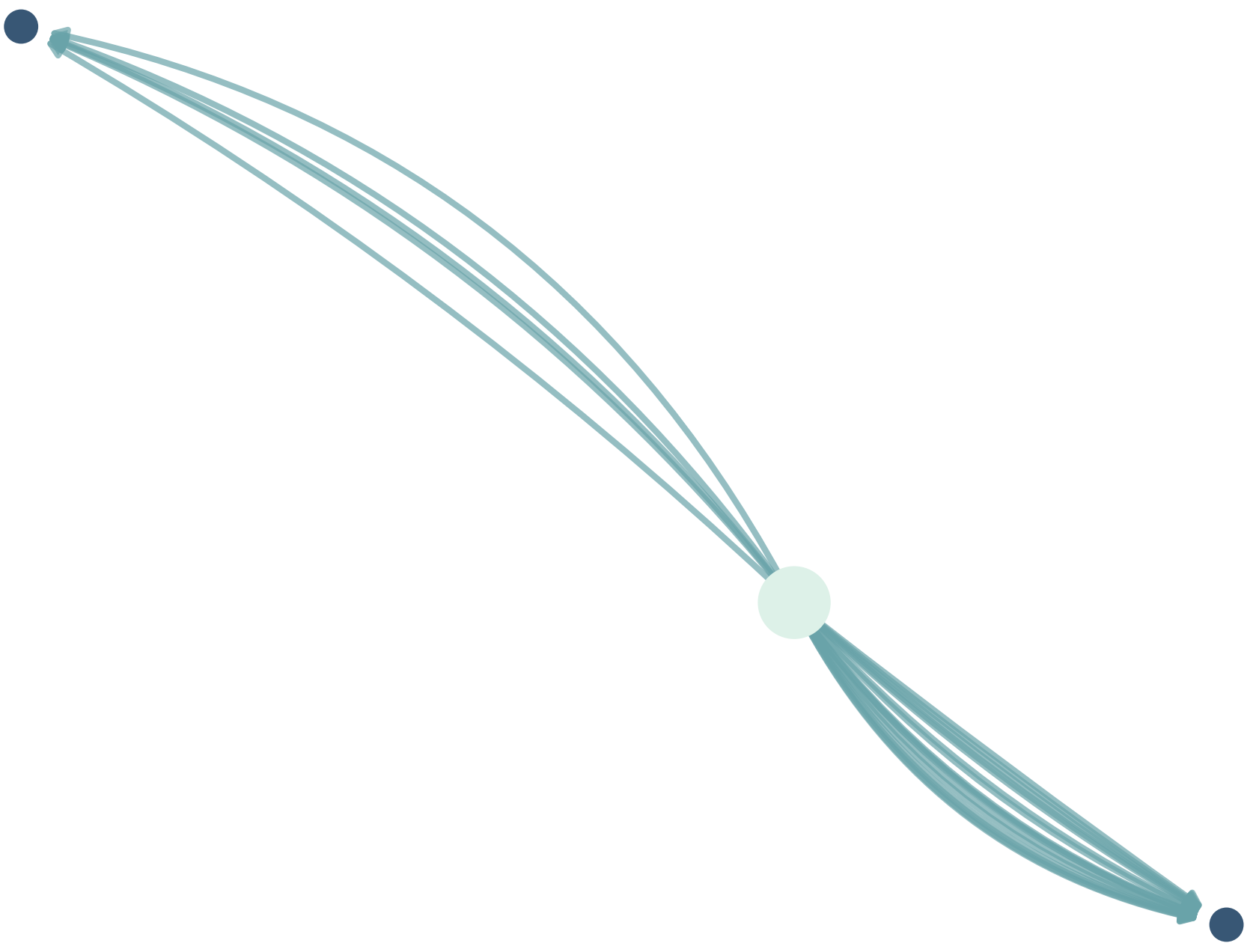}}\hspace{1pt}
\subfloat[\small{$(4\tau$, $5\tau)$}]{\includegraphics[width=.23\columnwidth]{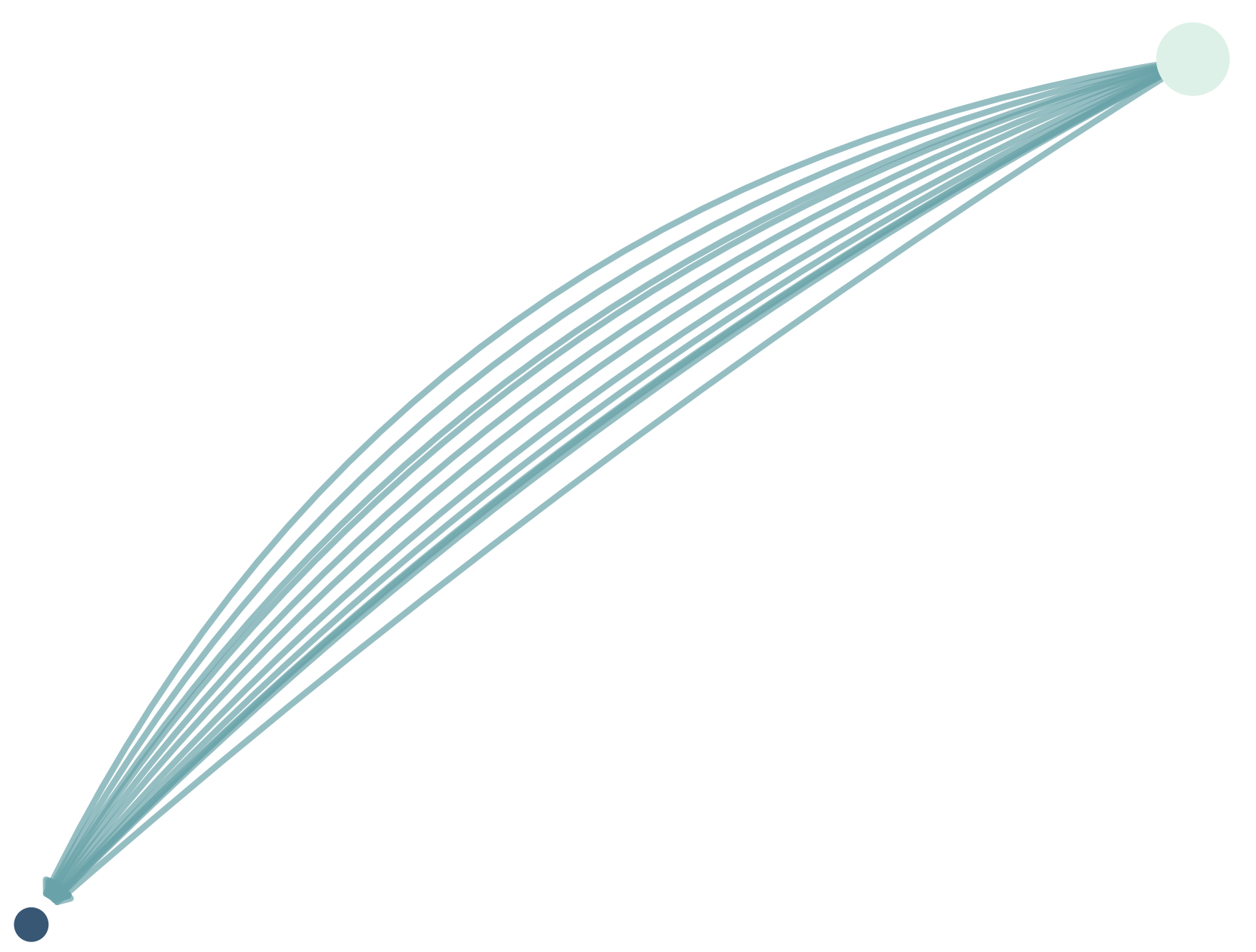}}\\
\subfloat[\small{$(5\tau$, $6\tau)$}]{\includegraphics[width=.24\columnwidth]{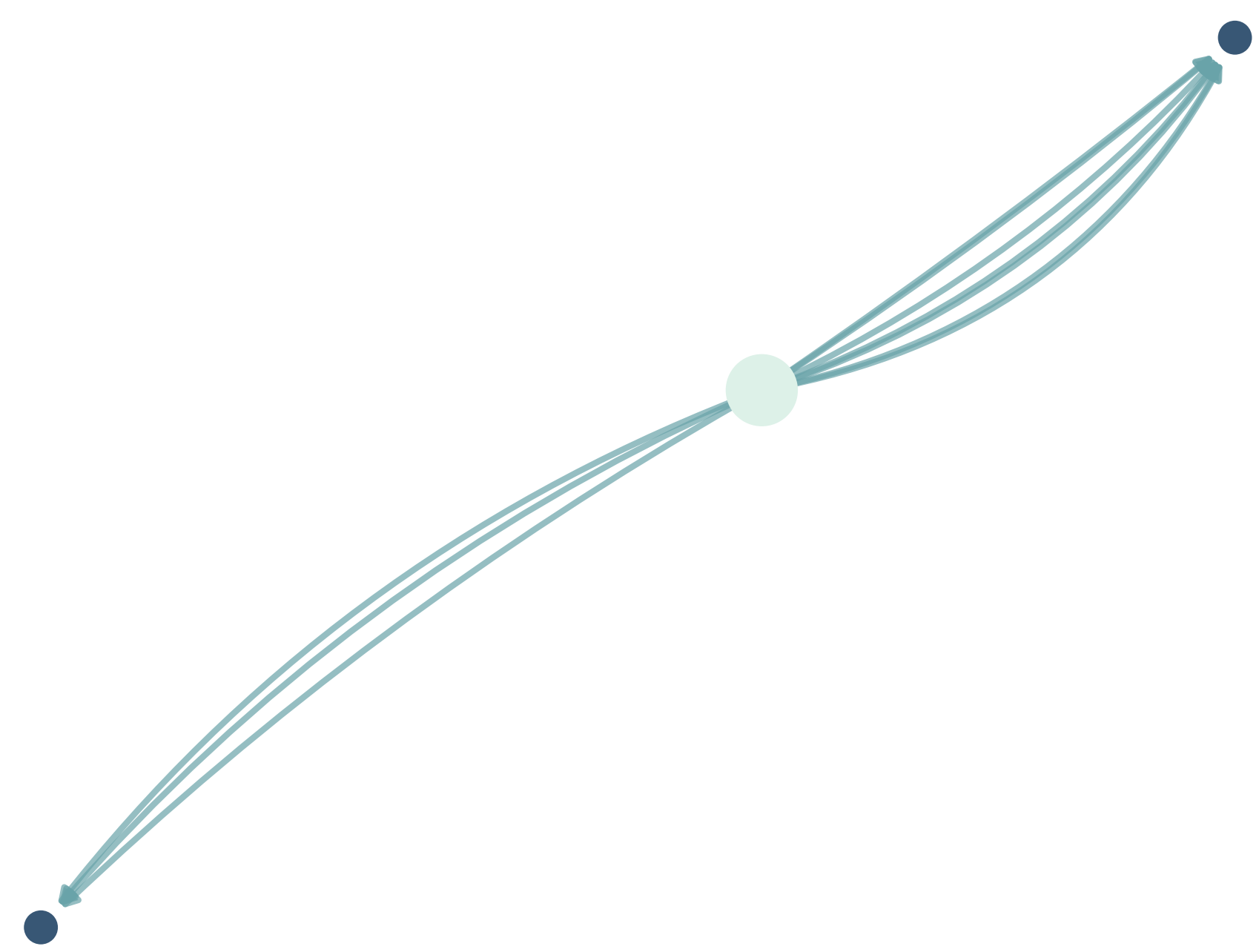}}\hspace{2pt}
\subfloat[\small{$(6\tau$, $7\tau)$}]{\includegraphics[width=.24\columnwidth]{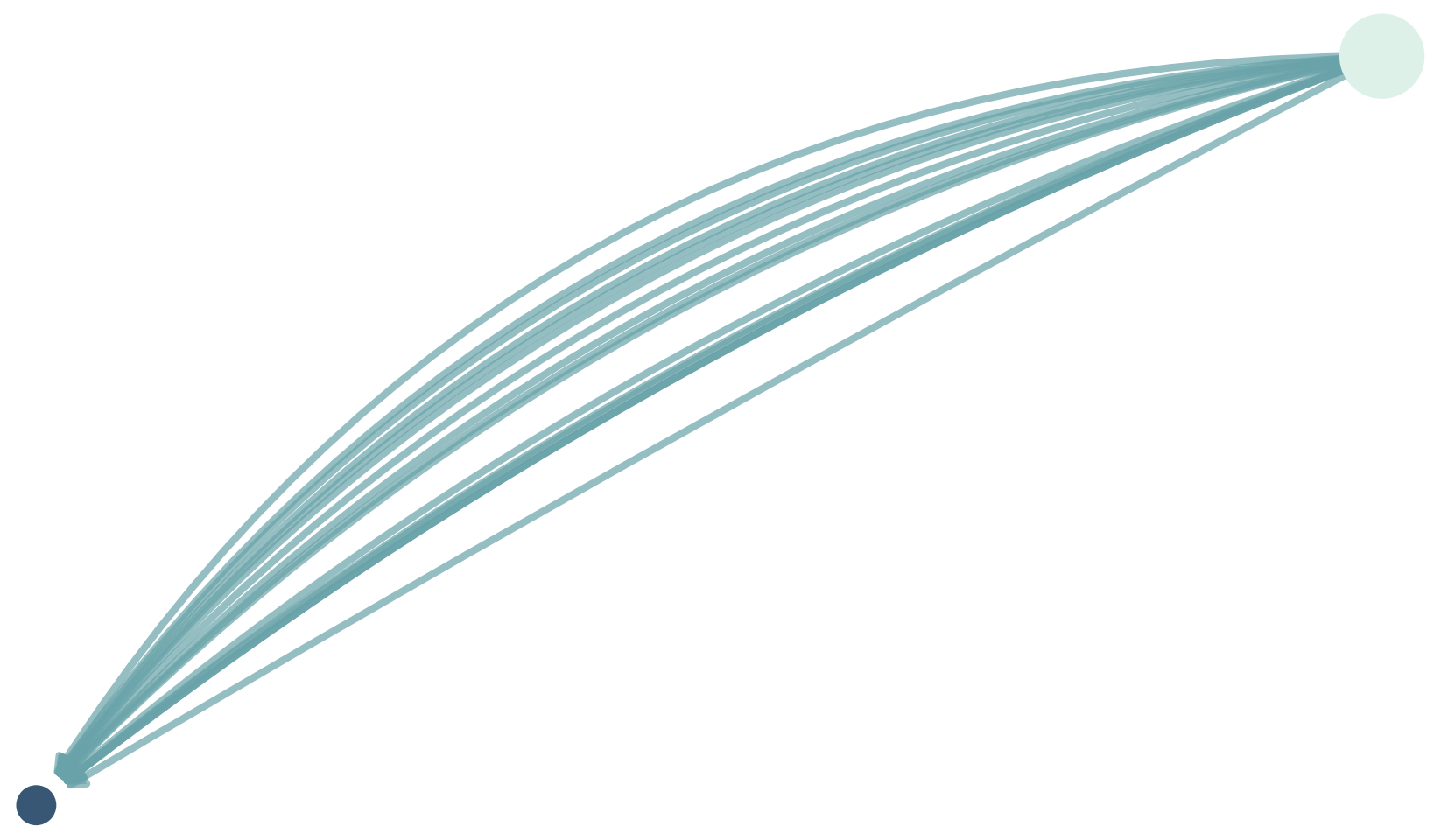}}\hspace{2pt}
\subfloat[\small{$(7\tau$, $8\tau)$}]{\includegraphics[width=.24\columnwidth]{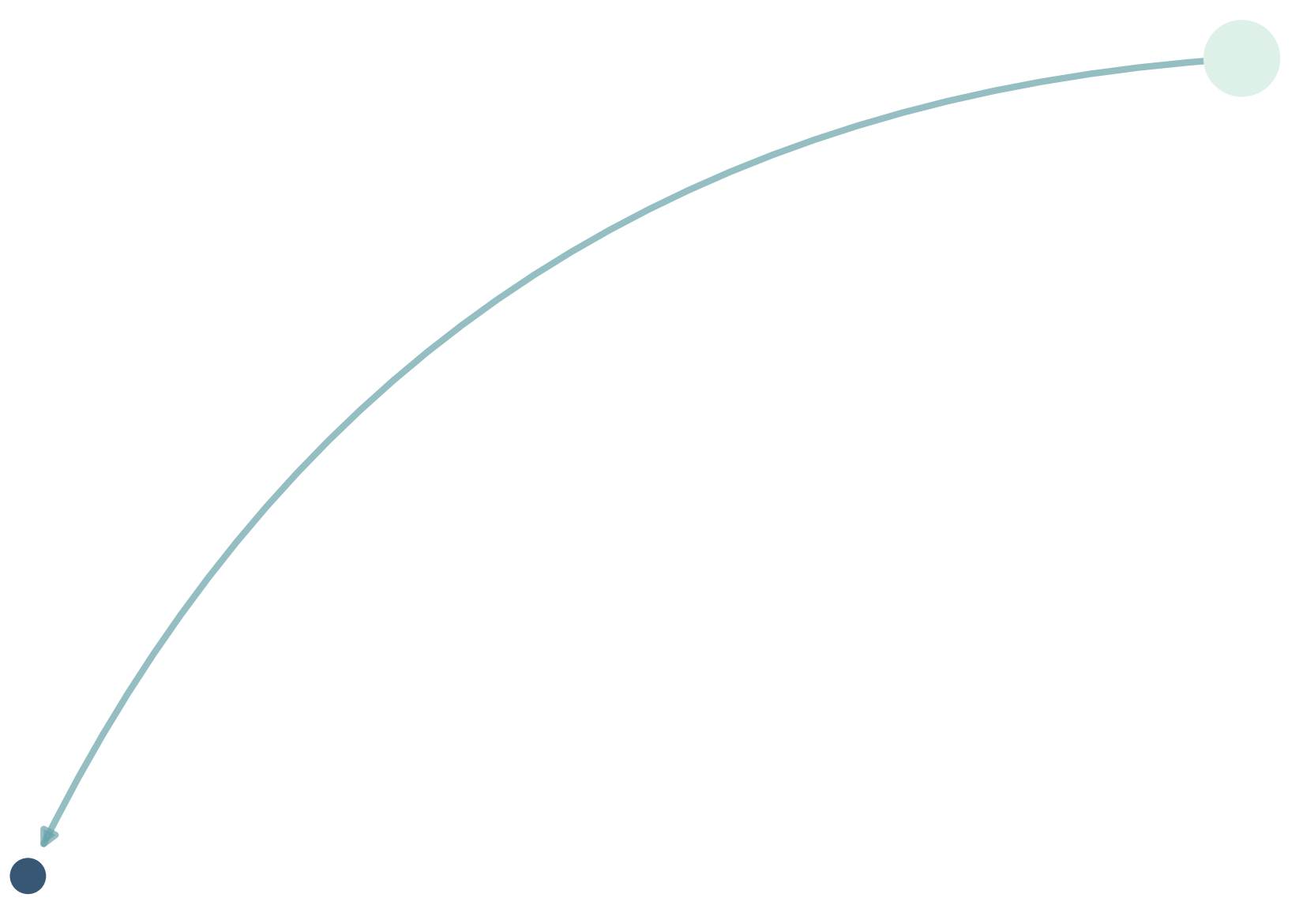}}\hspace{2pt}
\\
\vspace{-1ex}
\caption{A Dynamic Evolution Example of the Transaction Network of the Phishing Account.}
\label{fig:phishing_case_dynamicEvolution}
\end{figure}

Furthermore, traditional phishing accounts widely distribute phishing websites or scam information and steal the assets of some accounts (i.e., victims), without hiding behind the service providers. For the phishing scams shown in Fig. \ref{fig:phishing_case_dynamicEvolution}, we noticed that the number of accounts involved in each time interval is relatively small. However, the number of transactions between two nodes is relatively high.

\begin{figure}[h]
\centering
\vspace{-2ex}
\subfloat[\small{Original}]{\includegraphics[width=.49\columnwidth]{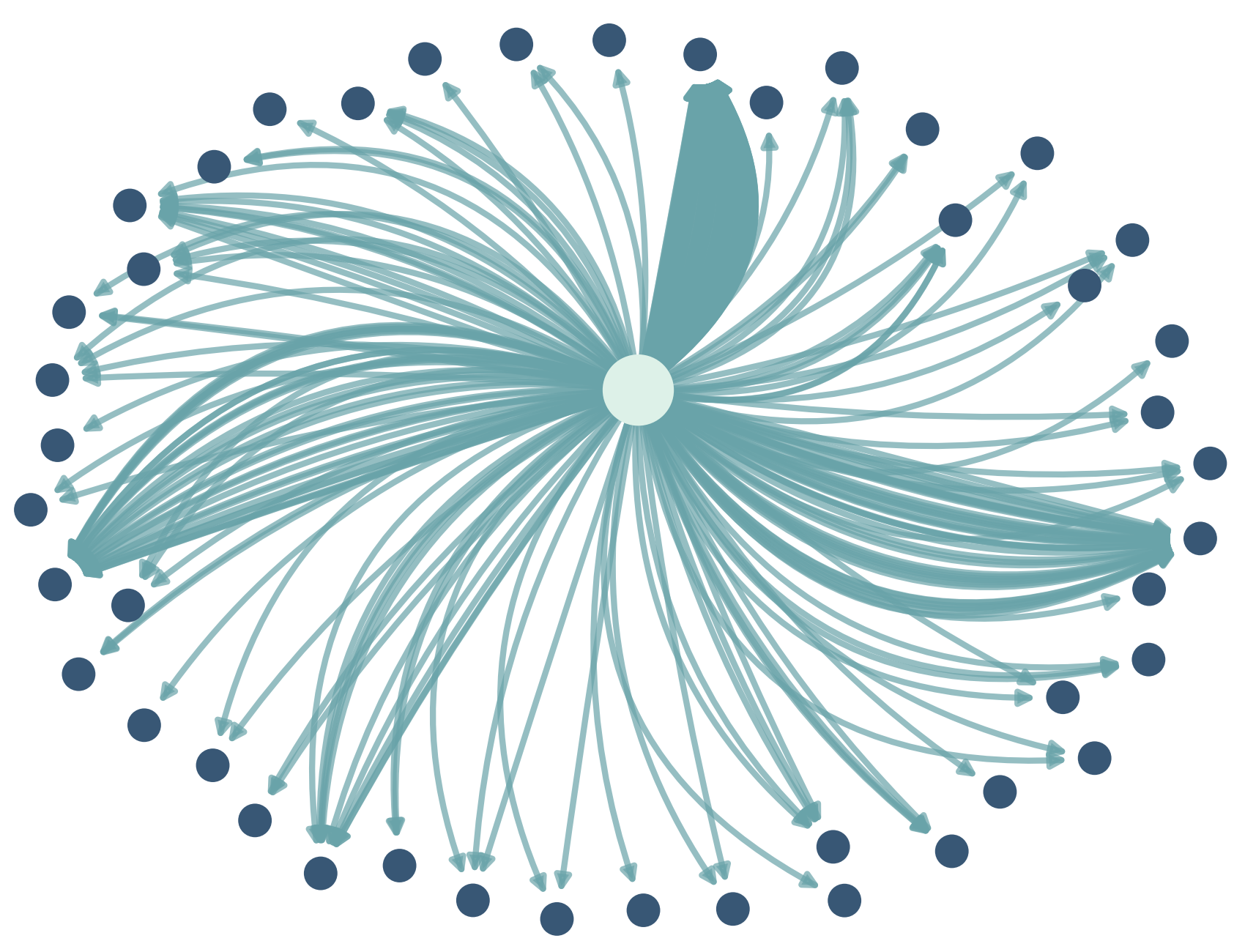}}\hspace{1pt}
\subfloat[\small{Detected Pattern}]{\includegraphics[width=.49\columnwidth]{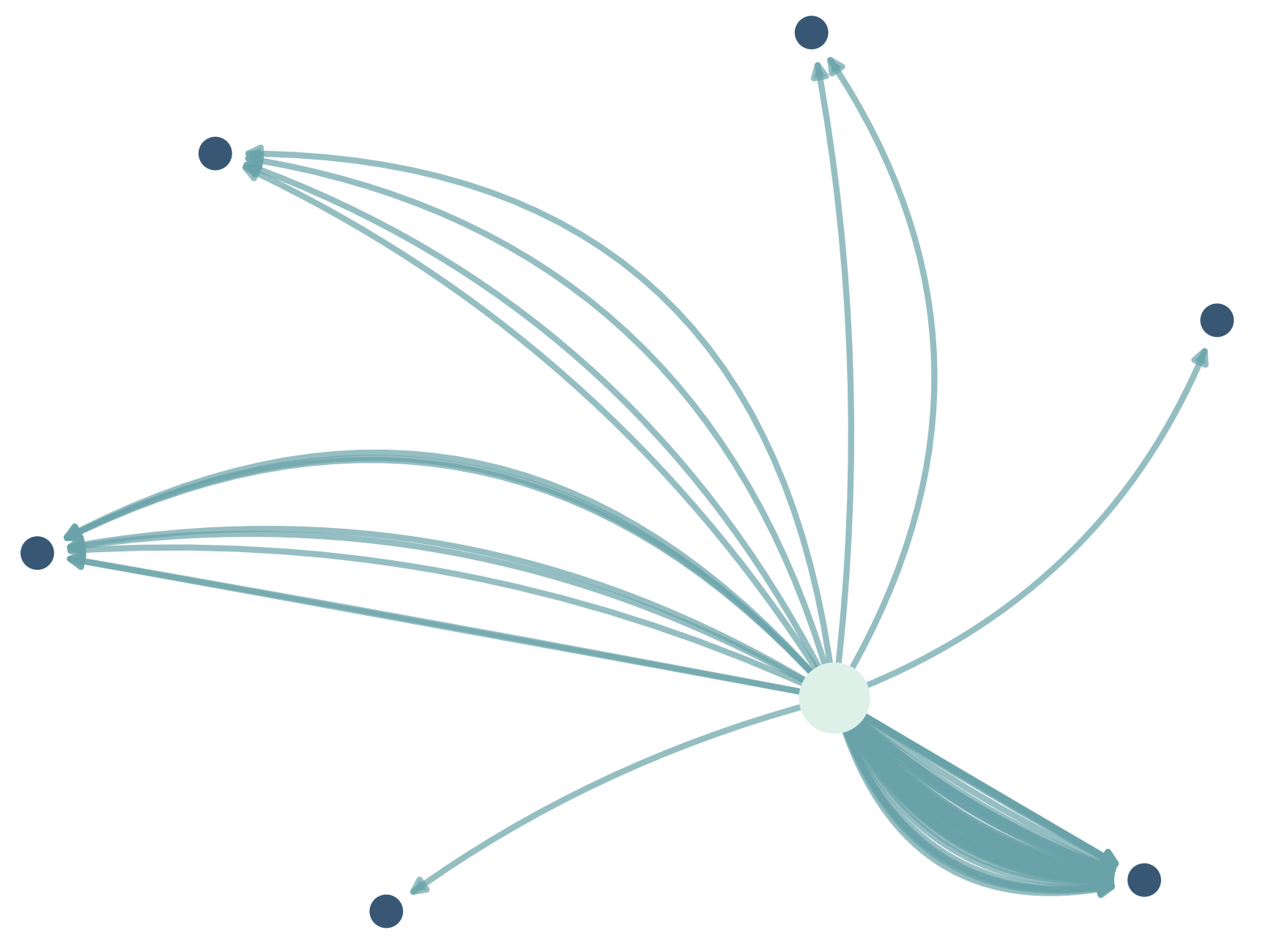}}\hspace{1pt}
\vspace{-1ex}
\caption{An Example of the Transaction Network of A Phishing Account, where \colorbox{mySourceNodecolor}{\phantom{}} represents the source account of the network, and \colorbox{myOtherNodecolor}{\phantom{}} means other connected accounts.}
\label{fig:phishing_case}
\vspace{-2ex}
\end{figure}

It shows that a phishing node initiates its activity by disseminating a transaction request. It is noticeable that there is a significant volume of transactions transpiring amidst nodes during the $(\tau,7\tau)$. This implies that the fraudulent accounts are selecting lucrative nodes and subsequently transferring their earnings during the $(7\tau,8\tau)$. In contrast to the means employed in Web3 scams, a phishing node does not revert to its prior condition following its initial engagement with a substantial number of accounts and subsequent profiting.

\subsection{Runtime Overhead}
The reason why ScamSweeper can detect the scam account in Fig. \ref{fig:Web3scams_case} is that it focuses not only on the graph structure but also on the sequential timestamp attributes. ScamSweeper takes the transaction graph as input, and we use the STRWalk to simplify the extensive transaction network. Moreover, as shown in Fig. \ref{fig:Web3scams_case}(b), ScamSweeper uses STRWalk to sample the original transactions, reducing the number of transactions from 399 to 134. 
The transaction pattern we detected, shown in Fig. \ref{fig:phishing_case}(b), is more simplified, and the pattern of 79 transaction numbers is detected from the 288 original transactions. 
It preserves their structural and temporal features while lowering the runtime overload, which helps to process large-scale transactions effectively.

\label{subsec:runtime_overhead}
\begin{table}[ht]
\centering
  \caption{The Runtime Results of Temporal Analysis with ScamSweeper on Different Graph Structures. \textbf{Spl} stands for the simplified graph obtained by sampling from STRWalk, and @5 means that the structure window size is 5.}
  \label{tab:runtime_overhead}
  \begin{tabular}{c c c c c} 
    \toprule
    \textbf{Graph Structure} & \textbf{Original} & \textbf{Spl@5} & \textbf{Spl@10} \\
    \midrule 
     \textbf{Time}  & 29.50 &  4.6673  &  5.0454  \\
     \textbf{F1-score} & 0.87 &  0.76   &   0.67  \\   
  \bottomrule 
\end{tabular}
\vspace{-2ex}
\end{table}

We perform a temporal analysis on 2,187 web3 scam accounts, which are randomly obtained from the dataset described at~\cref{subsubsec:setup}. We first get a simplified trading graph when STRWalk's window is set to 5 and 10. The original transactions are constructed as the initial transaction graph. Fig. \ref{fig:web3comparison} illustrates that ScamSweeper outperforms other methodologies when the structure windows are configured at 5 and 10. Thus, we present a detailed analysis of the time consumption associated with ScamSweeper across various structure windows. As shown in Table \ref{tab:runtime_overhead}, ScamSweeper takes 4.6673 and 5.0454 seconds on the simplified transaction graph with Spl@5 and Spl@10, while it takes more than 29.50 seconds on the initial transaction graph. The results show that the ScamSweeper with STRWalk can complete the detection with less time overhead.

\subsection{The Optimization of STRWalk}
\label{subsec:strwalk_optimation}
A series of procedures is conducted to assess the efficacy of STRWalk in transaction analysis. The findings, presented in Table \ref{tab:STRWalk_help}, provide a comparative analysis of the Node2Vec and DeepWalk methodologies applied to the original transaction network and the sampled network, which is constructed from the STRWalk algorithm, with \texttt{EPTransNet}~\cite{EPTransNet2024}. We can see that the sampling algorithm helped by STRWalk has a performance improvement. Under the F1 score, Node2Vec can improve from 0.26 to 0.79 with the advantage of around 203.85\%, and DeepWalk can improve from 0.21 to 0.80 with the advantage of about 280.95\%.

\begin{table}[ht]
\centering
  \caption{The Comparison Results on Phishing Nodes Detection. \dag\space refers to the utilization of STRWalk.}
  \label{tab:STRWalk_help}
  \begin{tabular}{c c c c c} 
    \toprule
    \textbf{Method} & \textbf{Precision} & \textbf{Recall} & \textbf{F1-score} \\
    \midrule
    Node2Vec~\cite{grover2016node2vec}            &  0.63  &  0.16  &  0.26  \\
    DeepWalk~\cite{perozzi2014deepwalk}           &  0.58  &  0.13  &  0.21  \\
    Node2Vec\dag~\cite{grover2016node2vec}            &  0.82  &  0.77  &  0.79  \\
    DeepWalk\dag~\cite{perozzi2014deepwalk}           &  0.82  &  0.79  &  0.80  \\
  \bottomrule 
\end{tabular}
\vspace{-2ex}
\end{table}

\subsection{The Temporal Setting for STRWalk}
\label{subsec:setting_strwalk}

\begin{figure}[ht]
\centering
\subfloat[ScamSweeper$^\dag$]{\includegraphics[width=.38\columnwidth]{Figures/embedding_ScamSweeper.pdf}}\hspace{1pt}
\subfloat[ScamSweeper$^\P$]{\includegraphics[width=.38\columnwidth]{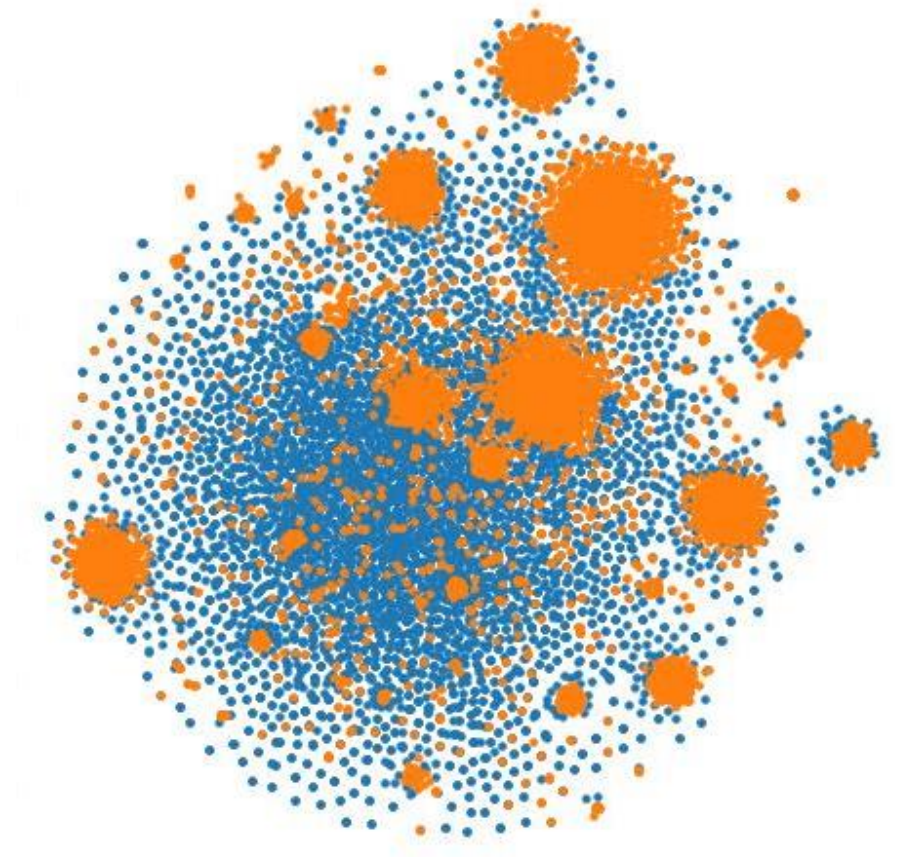}}\\
\caption{The Embedding Visualization of Node Feature Distribution by T-SNE, where \colorbox{orange}{\phantom{}} means malicious accounts, and \colorbox{blue}{\phantom{}} represents normal accounts. The $\dag$ and $\P$ refer to the STRWalk with Eq.\ref{eqa:temporal_walk_p} and Eq.\ref{eqa:temporal_walk_p_2}, respectively}
\label{fig:time_close_walks_Comparison}
\vspace{-2ex}
\end{figure}

In Eq.\ref{eqa:temporal_walk_p}, we subtract the minimum value from each edge to determine the probability weight, ultimately obtaining $P_i$ for random sampling in the first phase. It is based on the idea that transactions closer to their creation time are more important. To test this hypothesis, we perform an operation shown in Eq.\ref{eqa:temporal_walk_p_2} that contrasts with Eq.\ref{eqa:temporal_walk_p}, using the difference between the maximum time and the time of each edge.

\begin{equation}
\begin{aligned}
   P_i &= \frac{\mu(T[i])}{\sum_{j=1}^n \mu(T[j])},\\
   \mu(T[\ell]) &= \text{Max}(T) - T[\ell] + 1
\label{eqa:temporal_walk_p_2}
\end{aligned}
\end{equation}

After embedding the operation in Eq.\ref{eqa:temporal_walk_p_2} into STRWalk and training with ScamSweeper, we visualize malicious nodes and normal nodes using the t-SNE method in Fig. \ref{fig:time_close_walks_Comparison}.

In Fig. \ref{fig:time_close_walks_Comparison}, it is evident that although the majority of malicious nodes have been successfully clustered, there remains a notable overlap with normal nodes, presenting a significant challenge for effective segmentation.

\subsection{Threat to Validity}
\label{subsec:validity}

\noindent\textbf{Internal validity:} \textbf{(1)} In the web3 scams dataset (\cref{subsec:RQ1}), there are 8,736,430 nodes in our dataset, while there are only 3,125 malicious nodes. To balance it, we randomly sample an equal number of non-malicious nodes as negative samples. \textbf{(2)} In the phishing dataset (\cref{subsec:RQ1}), we collect 4,905 phishing and 636 normal nodes. We leverage the 636 normal nodes and EPTransNet, which is a subset dataset with 1,165 real phishing nodes. Overall, we construct a 1165:636 positive and negative dataset ratio to alleviate the imbalance.

\noindent\textbf{External Validity:} All the experimental datasets utilized in our study are collected from the real world, \textbf{(1)} Our dataset in RQ1 (\cref{subsec:RQ1}), where both phishing and normal nodes are obtained in Etherescan through the labels in XlabelCloud~\cite{xblock2024}. \textbf{(2)} As for the datasets of web3 scams, they are real scams that have been manually audited by Scamsniffer before collecting their transaction networks in Ethereum. Thus, these datasets are deemed highly suitable for evaluation. Note that all nodes present within the network can be detected concurrently, if necessary. However, due to imbalance issues, not all negative samples are subjected to testing.

\section{Conclusion}
\label{sec::conclusion}

Web3 services are widely distributed in the blockchain ecosystem, while many illegal accounts conceal malicious behavior when providing services. This paper proposes ScamSweeper, a detection model for detecting mimic accounts in web3 scams on Ethereum. We utilize the structure temporal random walk to sample the node network, which helps to alleviate the large-scale transaction network data. ScamSweeper can learn the sequence feature in the directed graph structure, exploring the dynamic evolution of time series. Through evaluating a real-world web3 scam dataset of transactions, ScamSweeper outperforms other methods by 15\% in weighted F1-score. Moreover, ScamSweeper has an advantage of 17.5\% in the F1-score on a phishing dataset.

\balance

\section{ACKNOWLEDGMENTS}
This work is sponsored by the National Natural Science Foundation of China (No.62362021, No.62402146, and No.62032025), and CCF-Tencent Rhino-Bird Open Research Fund (No.RAGR20230115).

\balance

\bibliographystyle{IEEEtran}
\normalem
\bibliography{ref}

\vfill

\end{document}